\documentclass{IEEEtran}
\usepackage{amsmath,amsfonts}
\usepackage{color}
\usepackage{array}
\usepackage{algorithmic}
\usepackage{textcomp}
\usepackage{stfloats}
\usepackage{url}
\usepackage{times}
\usepackage{verbatim}
\usepackage{graphicx}
\usepackage[nocompress]{cite}
\usepackage{longtable}
\usepackage[utf8]{inputenc}
\usepackage[T1]{fontenc}
\usepackage{lipsum}
\usepackage{multirow}
\usepackage{mathtools}
\usepackage{titlesec}
\usepackage[utf8]{inputenc}
\usepackage[linesnumbered,ruled,vlined]{algorithm2e}
\usepackage{amsmath, amssymb}
\begin{document}

\linespread{0.91}
\title{Study of Robust Resource Allocation in Cell-Free
Multiple-Antenna Networks \vspace{-0.35em}}

\author{Saeed Mashdour, André R. Flores, Rodrigo C. de Lamare  

\thanks{S. Mashdour, André R. Flores and  Rodrigo C. de Lamare are with Department of Electrical Engineering (DEE), Pontifical Catholic University of Rio de Janeiro, 22541-041 Rio de Janeiro, Brazil, Rodrigo C. de Lamare is also with the University of York, UK.  The emails are smashdour@gmail.com, \{andre.flores, delamare\}@puc-rio.br. This work was supported by CNPQ, CPQD, FAPERJ and FAPESP. }}

\maketitle

\begin{abstract}
Cell-free networks outperform cellular networks in many aspects, yet their efficiency is affected by imperfect channel state information (CSI). In order to address this issue, this work presents a robust resource allocation framework designed for the downlink of user-centric cell-free massive multi-input multi-output (CF-mMIMO) networks. This framework employs a sequential resource allocation strategy with a robust user scheduling algorithm designed to maximize the sum-rate of the network and two robust power allocation algorithms aimed at minimizing the mean square error, which are developed to mitigate the effects of imperfect CSI. An analysis of the proposed robust resource allocation problems is developed along with a study of their computational cost.  {Simulation results demonstrate the effectiveness of the proposed robust resource allocation algorithms, showing a performance improvement of up to 30\% compared to existing techniques.}

\end{abstract}

\section{Introduction}
Cell-free massive multi-input multi-output (CF-mMIMO) networks are a key innovation in wireless communications, which is characterized by the deployment of a large number of access points (APs) distributed over a wide area to serve multiple user equipment (UEs) simultaneously \cite{ngo2015cell} and  builds on the early work of massive multi-input multi-output (MIMO) systems \cite{marzetta,mmimo,wence}. This architecture enhances signal quality and capacity, mitigating the effects of fading and interference, common challenges in cellular systems \cite{interdonato2019ubiquitous}. By decentralizing the base station functionalities and using advanced signal processing techniques, CF-mMIMO networks promise significant improvements in data rates and reliability, representing a trend towards the next generation of wireless communication systems \cite{bjornson2020scalable}.

The architecture of CF-mMIMO networks is categorized into different frameworks, presenting approaches to managing the distributed nature of antennas and user equipment (UE) \cite{bjornson2020scalable, ammar2021user, mashdour2022enhanced}. In user-centric CF-mMIMO networks, each UE is served by a selected subset of APs, which could be chosen based on different criteria \cite{d2021improving, dong2019energy, d2020user, banerjee2022access, tds,jpba,mashdour2024clustering,rprs}. {While CF-mMIMO networks increase signal quality and system capacity, allocating resources in these networks is a crucial challenge to fully realize their potential \cite{demir2021foundations, nayebi2016performance, buzzi2017cell, zhang2024performance}.} Efficient resource allocation strategies are essential for balancing the load and ensuring fair service across the vast number of distributed APs and UEs. This involves sophisticated precoding \cite{siprec,gbd,wlbd,mbthp,robprec,bbprec,rcfprec,lrcc}, power control \cite{lrpa,rapa,rscf}, user scheduling, and bandwidth management to maximize network performance while minimizing interference and energy consumption \cite{d2021improving, mashdour2022multiuser, mashdour2023sequential}. 

Addressing these issues is fundamental to harnessing the true capabilities of CF-mMIMO networks, creating a foundation for a consistent and equitable wireless communication environment. In particular, precoding serves as a technique to pre-process the signals transmitted from multiple APs to mitigate interference and enhance the signal-to-interference-plus-noise ratio (SINR) for each UE, thereby optimizing the overall network throughput \cite{nayebi2017precoding}. Multiuser scheduling is mandatory to manage network effectively, especially in scenarios where the number of UEs exceeds the number of APs, such that the distribution of the available resources ensures that each UE is served by the most suitable subset of APs \cite{mashdour2022MMSE-Based, dimic2005downlink, ngo2015cell}. Power allocation strategies are also key to resource allocation, aiming to adjust the power levels across different APs to reach a tradeoff between network capacity maximization and energy efficiency \cite{van2016joint}.  

{The performance of CF-mMIMO networks is also influenced by factors like hardware impairments and the accuracy of channel state information (CSI) \cite{zhang2023much, zhang2023performance}.} Accurate CSI is crucial in CF-mMIMO networks, as it impacts essential tasks such as precoding, power control, and multiuser scheduling. Precise estimation of CSI ensures that signals are coherently combined at the intended UEs while minimizing interference, which is fundamental for network performance optimization. However, obtaining CSI is challenging due to factors such as channel estimation errors and feedback delays. Consequently, numerous studies have investigated practical models of imperfect CSI, accounting for estimation errors and delays to reflect real world conditions and develop algorithms that account for inaccuracies in CSI \cite{ interdonato2019ubiquitous, flores2022robust}. In CF-mMIMO networks with imperfect CSI, a key strategy is the development of robust resource allocation algorithms. The essence of these approaches lies in formulating resource allocation problems that can compensate for imperfect CSI. Such robust designs are crucial for mitigating the detrimental effects of imperfect CSI on network performance \cite{ bjornson2015optimal}.

\subsection{Prior and related works}

{The study of robust transmit processing techniques in wireless networks in the presence of imperfect CSI scenarios has emerged as an important topic for both cellular and cell-free network frameworks. 
The work of \cite{rong2005robust} investigated robust linear receiver techniques that enhance joint space-time decoding and interference rejection using a worst-case optimization approach, where generalizations of minimum variance methods are designed to tackle the challenges of imperfect CSI at the receiver.}

{In \cite{bashar2018robust} the challenge of optimizing user scheduling in the uplink of massive MIMO systems is considered with the aim of decreasing the channel estimation overhead, using the COST 2100 channel model. This approach reduced the need for accurate CSI and ensures the effectiveness of user scheduling. The work in
\cite{vaezy2018energy} examines the design of precoders for a multiuser MIMO (MU-MIMO) system under two distinct conditions: firstly, where imperfect CSI is present with known channel error statistics, and secondly, where there is a norm-bounded error without statistical characterization. This study introduced an energy-efficient precoder design for MU-MIMO systems crafted to enhance robustness against imperfect CSI.}

{The study of \cite{chang2017energy} investigated the joint optimization of antenna selection, power allocation, and time allocation to maximize the system energy efficiency (EE) of a wireless power transfer (WPT) enabled multiuser massive MIMO system. With users empowered solely by WPT in the downlink, the work accounts for practical constraints, including imperfect CSI. 
In \cite{wang2009worst}, robust transmit strategies have been presented for MIMO systems to address imperfect CSI  using a worst-case deterministic model. It focuses on maximizing the received signal-to-noise ratio (SNR) or minimizing the Chernoff bound on error probability, aiming to solve max-min and quality of service (QoS) problems efficiently by transforming them into convex problems or semidefinite programs (SDPs). 
The work in \cite{palhares2021robust} investigated the downlink of a CF-mMIMO network with single-antenna APs and UEs, focusing on robustness against imperfect CSI. It introduced an iterative robust MMSE (RMMSE) precoder incorporating generalized loading to mitigate interference caused by imperfect CSI along with optimal and uniform power allocation schemes.} 

\subsection{Contributions}

{In this work, we propose a robust resource allocation framework, which employs a sequential strategy for multiuser scheduling and power allocation, to mitigate the effects of imperfect CSI in the downlink of CF-mMIMO networks \cite{rspa}. 
In particular, we adopt linear and Minimum Mean Square Error (MMSE) or Zero-Forcing (ZF) precoders as the basic transmit processing filters for multiuser interference mitigation. Furthermore, a robust multiuser scheduling algorithm denoted robust clustered enhanced subset greedy (RC-ESG) algorithm that implements worst-case robust optimization techniques is developed, which accounts for imperfect CSI and ensures that user scheduling decisions are resilient to estimation errors. Moreover, we propose two robust power allocation algorithms to enhance the network's information rates and robustness in the presence of imperfect CSI. The first technique integrates channel estimation errors directly into the optimization problem and employs a robust gradient descent power allocation (RGDPA) algorithm, which enables a cost-effective power allocation. The second technique, rooted in worst-case robust optimization (WRGDPA), optimizes power allocation by preparing for the most unfavorable scenarios of channel uncertainty. 
We also carry out a convexity analysis of the proposed robust approaches along with a study of their computational costs. }

\vspace{-2mm}
{\subsection{Paper Outline and Notation}}
The rest of this paper is organized as follows: Section \ref{SYS.mod} provides a model for UCCF networks, including their sum-rate, and methods for resource allocation. Section \ref{Robust.PA} presents a novel robust multiuser scheduling approach along two innovative robust power allocation methods, one incorporating the channel estimation error matrix and the other based on a worst-case scenario approach. Section \ref{Analysis} presents an analysis of solutions to various optimization problems in the proposed robust user scheduling method, the convexity of the objective functions specific to the proposed robust power allocation techniques, and compares the computational complexity of the proposed algorithms. Section \ref{Simul} shows and discusses the results of the simulations conducted to assess the algorithms. Finally, Section \ref{conclud} concludes the paper.

We use the following notations. $\mathbf{I}_{n}$ is an identity matrix of dimensions $n\times n$. The complex normal distribution is denoted by $\mathcal{CN}\left (.,. \right )$. The superscripts $^{T}$, $^{\ast}$, and $^{H}$ represent the operations of transpose, complex conjugate, and Hermitian (conjugate transpose), respectively. The symbol $\mathcal{A}\cup \mathcal{B}$ indicates the union of sets $\mathcal{A}$ and $\mathcal{B}$, while $\mathcal{A}\setminus \mathcal{B}$ is the set difference, indicating elements in $\mathcal{A}$ not in $\mathcal{B}$. Complex matrices with dimensions $M\times N$ are represented by $\mathbb{C}^{M\times N}$. The trace of a matrix is indicated by $Tr\left (.\right )$. The function $\textup{diag}\left ( \mathbf{x} \right )$ converts the vector $\mathbf{x}$ into a diagonal matrix, and $\textup{diag}\left ( \mathbf{X} \right )$ extracts the diagonal elements from the matrix $\mathbf{X}$. The expectation operator is denoted by $\mathbb{E}\left [ . \right ]$. Finally, $[\mathbf{X}]_{m,n}$ denotes the element located in the $m$th row and $n$th column in $\mathbf{X}$.

\section{System model and problem statement} \label{SYS.mod}

{We consider the downlink of a CF-mMIMO network architecture consisting of $L$ APs, each equipped with an array of $N$ antenna elements, which serve $K$ single-antenna UEs.} The considered network structure is distinguished by a user-centric approach for selecting APs to enhance overall network performance. A key aspect of this architecture is that the total number of UEs, $K$, significantly exceeds the total number of AP antennas $M=LN$ ($K\gg M$). This implies the need for effective user scheduling to ensure efficient system operation. Thus, at each resource block transmission, the network schedules a subset of $n \leq M$ UEs. Additionally, the network can also allocate the powers of the $n$ scheduled users to enhance the sum-rates. By customizing the AP-UE associations and scheduling decisions based on real-time channel conditions and network demands, our objective is to design a system that enhances network performance by maximizing the sum-rates.

\subsection{User-Centric CF (UCCF) Network Architecture}

{The channel matrix representing the propagation effects between APs and UEs is given by $\mathbf{G}=\hat{\mathbf{G}}+\tilde{\mathbf{G}}\in \mathbb{C}^{LN\times n}$, where $\hat{\mathbf{G}}$ is the channel estimation matrix and $\tilde{\mathbf{G}}$ is the channel estimation error. The channel estimation error $\tilde{\mathbf{G}}$ is modeled as a complex Gaussian random variable with zero mean and known variance to reflect the stochastic nature of estimation uncertainties \cite{nayebi2017precoding}.} {Accordingly, $g_{mk} = \left[ \mathbf{G} \right]_{m,k}$ refers to the channel coefficient between the $m$th AP antenna (one of the $M = LN$ AP antennas) and the $k$th UE, representing the element of the channel matrix $\mathbf{G}$ located at the $m$th row and the $k$th column, and is modeled as follows:} 
\begin{equation} \label{eq.gI}
\begin{split}
    g_{{mk}} = & \hat{g}_{mk}+\tilde{g}_{mk}\\
    =&\sqrt{1-\alpha}\sqrt{\beta_{mk}}h_{mk} + \sqrt{\alpha}\sqrt{\beta_{mk}}\tilde{h}_{mk},
\end{split}
\end{equation}
where $\hat{g}_{mk}$ is the channel coefficient estimate, $\tilde{g}_{mk}$ is the channel coefficient estimation error, \( 0<\alpha<1 \) is a CSI imperfection parameter that balances the contribution of the small scale channel estimation coefficient \( h_{mk} \) and its estimation error \( \tilde{h}_{mk} \). {Note that the large-scale fading (LSF) coefficient $\beta_{mk}$ is assumed to be known, as it can be reliably estimated and tracked due to its slow variation, consistent with the findings in \cite{nayebi2017precoding}.} Furthermore, the variable \( h_{mk} \) which is modeled by independent and identically distributed (i.i.d.) random variables (RVs), is constant within a coherence interval while varying independently across different coherence intervals~\cite{ngo2017cell}, and follows a complex Gaussian distribution with zero mean and unit variance. Similarly, \( \tilde{h}_{mk} \) has a complex Gaussian distribution with zero mean and unit variance \cite{flores2023clustered}, and is independent of \( h_{mk} \) as supported by \cite{nayebi2017precoding, mishra2022rate}. In our system model, we employed linear ZF and MMSE precoders, denoted by \( \mathbf{P} \in \mathbb{C}^{LN\times n}\), to optimize the signal reception quality. Thus, the downlink signal is modeled by
\begin{equation}\label{rec.sig}
\begin{split}
\mathbf{y}=&\sqrt{\rho_{f}}\mathbf{G}^T\mathbf{P}\mathbf{x}+\mathbf{w}\\
=&\sqrt{\rho_{f}}\hat{\mathbf{G}}^T\mathbf{P}\mathbf{x}+\sqrt{\rho_{f}}\tilde{\mathbf{G}}^T\mathbf{P}\mathbf{x}+\mathbf{w},
\end{split}
\end{equation}
where $\rho_{f}$ is the maximum transmitted power of each antenna, $\mathbf{x} = \left[ x_{1}, \cdots, x_{n} \right]^{T}$ denotes the symbol vector, characterized by a zero mean and distributed as $\mathbf{x} \sim \mathcal{CN}(\mathbf{0}, \mathbf{I}_{n})$. Additionally, $\mathbf{w} = \left[ w_{1}, \cdots, w_{n} \right]^{T}$ represents the additive noise vector, following the distribution $\mathbf{w} \sim \mathcal{CN}(\mathbf{0}, \sigma_{w}^{2}\mathbf{I}_{n})$. In this model, we adopt Gaussian signaling and assume that the elements of $\mathbf{x}$ are statistically independent, and also independent from all noise components and channel coefficients.

\begin{figure}
    \centering
\includegraphics[width=.725\linewidth]{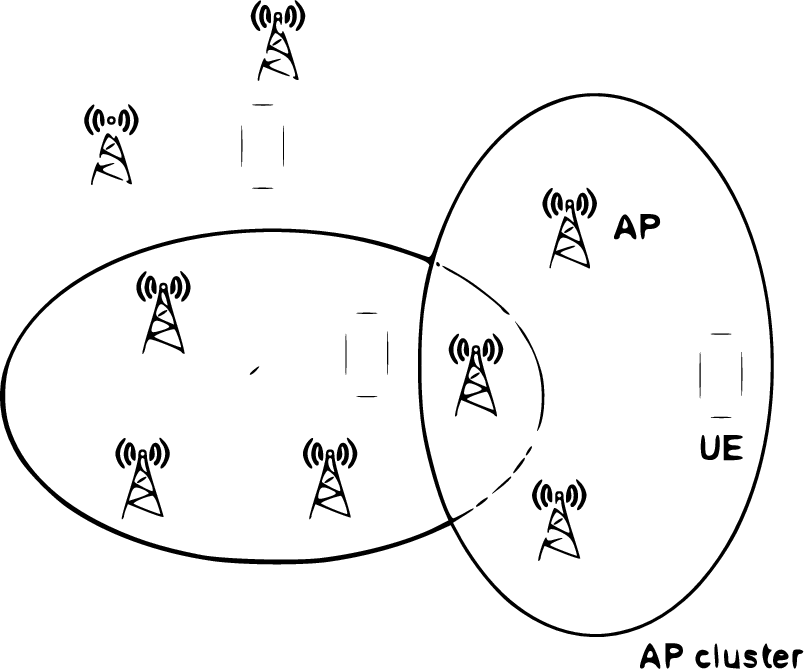}
    \vspace{-1em}
    \caption{Illustration of a user-centric cell-free network and the clustered APs}
    \label{model}
\end{figure}

In the UCCF network structure depicted in Fig.~\ref{model}, we cluster the APs supporting each 
UE based on the LSF criterion. Thus, the channel matrix in UCCF is defined as $\mathbf{G}_{a}=\left [ \mathbf{g}_{a1},\cdots , \mathbf{g}_{an}\right ]\in \mathbb{C}^{LN\times n}$ in which $\mathbf{g}_{ak}=\mathbf{A}_{k}\mathbf{g}_{k} \in \mathbb{C}^{{LN}\times 1}$ is the channel vector to the UE $k$, $k \in \left \{ 1,\cdots ,n \right \}$, $\mathbf{g}_{k}$ showing the CF channel vector of the UE $k$ and $\mathbf{A}_{k}$ is a diagonal matrix with the entries defined as follows
\begin{equation} \label{akl}
    a_{kl}=\left\{\begin{matrix}
1 & \textrm{if} \  l\in U_{k}\\ 
0 & \textrm{if} \  l\notin  U_{k}
\end{matrix}\right.
, l\in \left \{ 1,\cdots ,LN \right \},
\end{equation}
The LSF based clustering approach is based on the macroscopic propagation characteristics, such as path loss and shadowing, to allocate potential APs to the UEs.

In the LSF-based AP clustering approach, an AP $m$ is found to be suitable for a user $k$ if its average channel gain exceeds a predefined threshold $\lambda_{lsf}$, satisfying $\beta_{mk} \ge \lambda_{lsf}$, where $\beta_{mk}$ is the large-scale fading coefficient described before. In the case that no AP meets this criterion for a particular user, the AP with the largest average channel gain is chosen as the sole member of the corresponding subset \cite{ammar2021downlink}. Consequently, the assembling of the chosen APs for user $k$ is given by
\vspace{-2mm}
\begin{equation} \label{eq:CF-sig}
    U_{k} = \left\{ m : \beta_{mk} \ge \lambda_{lsf} \right\} \cup \left\{ \arg \max_{m} \beta_{mk} \right\},
\end{equation}
where the threshold $\lambda_{lsf}$ is determined by
\vspace{-2mm}
\begin{equation} \label{eq:alphalsf}
    \lambda_{lsf} = \frac{1}{LN \cdot K} \sum_{m=1}^{LN} \sum_{k=1}^{K} \beta_{mk}.
\end{equation}
{In this user-centric clustering approach, the set of APs \( U_{k} \) serving UE \( k \) is determined based on the LSF criterion. Specifically, the set \( U_{k} \) includes all APs where the LSF coefficient \( \beta_{mk} \) exceeds a predefined threshold \( \lambda_{lsf} \), ensuring that only APs with sufficiently strong signals are selected. Therefore, the AP with the maximum LSF coefficient for each UE is always included in the set \( U_{k} \) \cite{ammar2021downlink}.}  Moreover, the linear precoder matrix in the UCCF network is defined as $\mathbf{P}_{a}\in \mathbb{C}^{LN\times n}$ as a function of $\mathbf{G}_{a}$. Thus, using (\ref{rec.sig}), we can write the downlink received signal for UCCF networks as
\begin{equation} \label{UCCF.sig}
\begin{split}
    \mathbf{y}_a=&\sqrt{\rho _{f}}\mathbf{G}_{a}^T\mathbf{P}_a\mathbf{x}+\mathbf{w}\\
=&\sqrt{\rho_{f}}\hat{\mathbf{G}}_{a}^T\mathbf{P}_{a}\mathbf{x}+\sqrt{\rho_{f}}\tilde{\mathbf{G}}_{a}^T\mathbf{P}_{a}\mathbf{x}+\mathbf{w}.
\end{split}
\end{equation}
The signal received, as described in (\ref{UCCF.sig}), is based on the sparse channel matrix $\mathbf{G}_{a}$. In this matrix, for $k$th column, $k\in \left \{ 1, \cdots , n  \right \}$, only the elements corresponding to the APs that are determined to support UE $k$ are non-zero, while all other elements are zero.

\subsection{UCCF Sum-Rate Expression}

We assume that the received signal has an error component associated with imperfect CSI, as characterized by the received signal model in (\ref{UCCF.sig}). Nevertheless, the channel model and the associated error magnitude are constructed in such a manner that the channel is neither regarded as perfectly known nor entirely unreliable. Consequently, under the assumption of Gaussian signaling, and statistically independent $\mathbf{x}$ and $\mathbf{w}$, the expression for the sum-rate is given by the following equation as proved in Appendix \ref{sum-rate}:

\begin{equation}\label{eq:RCF}
    {SR = \log_2\left(\det\left[\mathbf{R}_{UC_{\tilde{\mathbf{G}}_a}} + \mathbf{I}_K\right]\right),}
\end{equation}
{where the matrix $\mathbf{R}_{UC_{\tilde{\mathbf{G}}_a}}$ is:}
\vspace{-2mm}

\begin{equation}\label{eq:RCF_1}
    {\mathbf{R}_{UC_{\tilde{\mathbf{G}}_a}} =} {\rho_f \hat{\mathbf{G}}_a^T \mathbf{P}_a \mathbf{P}_a^H \hat{\mathbf{G}}_a^* \left( \mathbf{R}_{\tilde{\mathbf{G}}_a} \right)^{-1},}
\end{equation}

\begin{equation}
    {\mathbf{R}_{\tilde{\mathbf{G}}_a} = \mathbb{E}_{\tilde{\mathbf{G}}_a} \left[ \rho_f \tilde{\mathbf{G}}_a^T \mathbf{P}_a \mathbf{P}_a^H \tilde{\mathbf{G}}_a^* \right] + \sigma_w^2 \mathbf{I}_K.}
\end{equation}
{We acknowledge that for the channel model considered in (\ref{eq.gI}), longer pilot sequences may improve channel estimation accuracy and increase the sum-rate \cite{nayebi2017precoding}, but they also introduce more signaling overhead, which could reduce the sum-rate. However, since our focus is on the downlink, we leave the investigation of this impact on the uplink for future works.}

\subsection{Resource Allocation in UCCF Networks}

Given the constraint that the number of UEs $K$  exceeds the total number of AP antennas $M$, user scheduling becomes necessary which ensures that a manageable and efficient subset of UEs is served at any given time. Defining the user scheduling task, we formulate the following optimization problem as described in \cite{mashdour2022enhanced}:
\begin{equation} \label{rop1}
    \begin{aligned}
& \underset{\mathcal{S}_{n}}{\text{max}}~SR\left ( \mathcal{S}_{n} \right ) \\
& \text{subject to} \ \left \| \mathbf{P}_a\left ( \mathcal{S}_{n} \right ) \right \|_{F}^{2}\leq P,
\end{aligned}
\end{equation}
where $\mathcal{S}_{n}$ is the set of $n$ UEs to be scheduled, and $P$ is the upper limit of the signal covariance matrix ${Tr}\left ( \mathbf{C}_{\mathbf{x}}\right )\leq P$ and $\mathbf{P}_a\left ( \mathcal{S}_{n} \right )\in \mathbb{C}^{LN\times {n}}$ is the precoding matrix of the scheduled users.
{The sum-rate given in (\ref{eq:RCF}) is a complex expression that becomes more involved when incorporating the MMSE precoder, which involves matrix inversions and other mathematical operations, making the optimization problem in \eqref{rop1} highly challenging. Therefore, to solve the optimization problem in \eqref{rop1} we resorted to a greedy method. }

{For power allocation, we adopt a gradient method to solve the optimization problem in \eqref{rop1} using an equivalent MSE minimization approach. The rationale behind this is that minimizing the MSE maximizes the signal to interference-plus-noise ratio (SINR), which corresponds to maximizing the sum-rate \cite{verdu1998multiuser} under Gaussian signaling. This equivalence holds for single-stream systems  and in more general multi-stream scenarios with interference coupling, this equivalence becomes an approximation. However, as highlighted in \cite{palomar2003joint} and \cite{ngo2013energy}, the relationship between SINR and MMSE is well-established under Gaussian signaling and is widely adopted as a tractable surrogate for capacity maximization. By focusing on a more manageable MSE minimization for power allocation, we develop simpler algorithms that approximate sum-rate maximization effectively.}

In this regard, we first examine (\ref{UCCF.sig}), which includes a desired term as well as terms associated with imperfect CSI and noise. By integrating power allocation into the received signal, we arrive at the signal described by
\begin{equation} \label{yc2}
\begin{split}
\mathbf{y}_a&=\sqrt{\rho _{f}}\hat{\mathbf{G}}_a^T\mathbf{W}\mathbf{D}\mathbf{x}+\sqrt{\rho _{f}}\tilde{\mathbf{G}}_a^T\mathbf{P}_{a}\mathbf{x}
+\mathbf{w},
\end{split}
\end{equation}
where the precoding matrix in the desired term can be rewritten as $\mathbf{P}_a=\mathbf{W}\mathbf{D}$ such that $\mathbf{W}\in\mathbb{C}^{NL \times {n}}$ is the normalized linear precoder matrix and $\mathbf{D}\in\mathbb{C}^{{n} \times {n}}$ is the power allocation matrix described by
\begin{equation}
\mathbf{D}=\begin{bmatrix}
\sqrt{p_{1}} & 0 & \cdots  & 0\\ 
 0& \sqrt{p_{2}} & \cdots &0 \\ 
 \vdots & \vdots  &\cdots   & \vdots \\ 
 0&0  &\cdots   & \sqrt{p_{{n}} }
\end{bmatrix}=\textup{diag}\left ( \mathbf{d} \right ),
\end{equation}
where $\mathbf{d}
=\left [ \sqrt{p_{1}} \ \sqrt{p_{2}} \ \cdots \  \sqrt{p_{{n}}} \right ]^T$.

In order to perform power allocation, we use the MSE between the transmitted signal and the received signal as the objective function and consider the optimization problem \cite{mashdour2022MMSE-Based}:
\begin{equation} \label{optmse.1}
\begin{aligned}
& {\underset{\mathbf{d}}{\text{min}}~\mathbb{E}\left [ \varepsilon  \right ]} \\
& {\text{subject to}\ \left \| \mathbf{W} \textup{diag}\left ( \mathbf{d} \right ) \right \|^{2}\leq P,} 
\end{aligned}
\end{equation}
where the error is
\begin{equation}\label{err.def}
    \varepsilon =\left \| \mathbf{x}-\mathbf{y}_a \right \|^{2}
\end{equation}
{and a non-negativity constraint on the power levels, i.e., ${\mathbf{d} \succeq 0 }$ was found to be unnecessary as, in our simulation studies, the proposed algorithms resulted in non-negative power levels.}

When examining the specified optimization problems, it becomes evident that imperfect CSI can lead to suboptimal decisions and performance degradation. This necessitates a robust optimization approach capable of mitigating the uncertainties introduced by CSI errors.

{The network's scalability can be enhanced by first performing AP selection, ensuring each AP only serves a limited number of UEs, reducing computational load and fronthaul requirements. This approach, combined with linear MMSE precoder and user scheduling before power allocation, aligns with the findings in \cite{bjornson2020scalable} and preserves scalability.}

\section{Robust Resource Allocation} \label{Robust.PA}

\begin{figure}
	\centering
\includegraphics[width=0.91\linewidth]{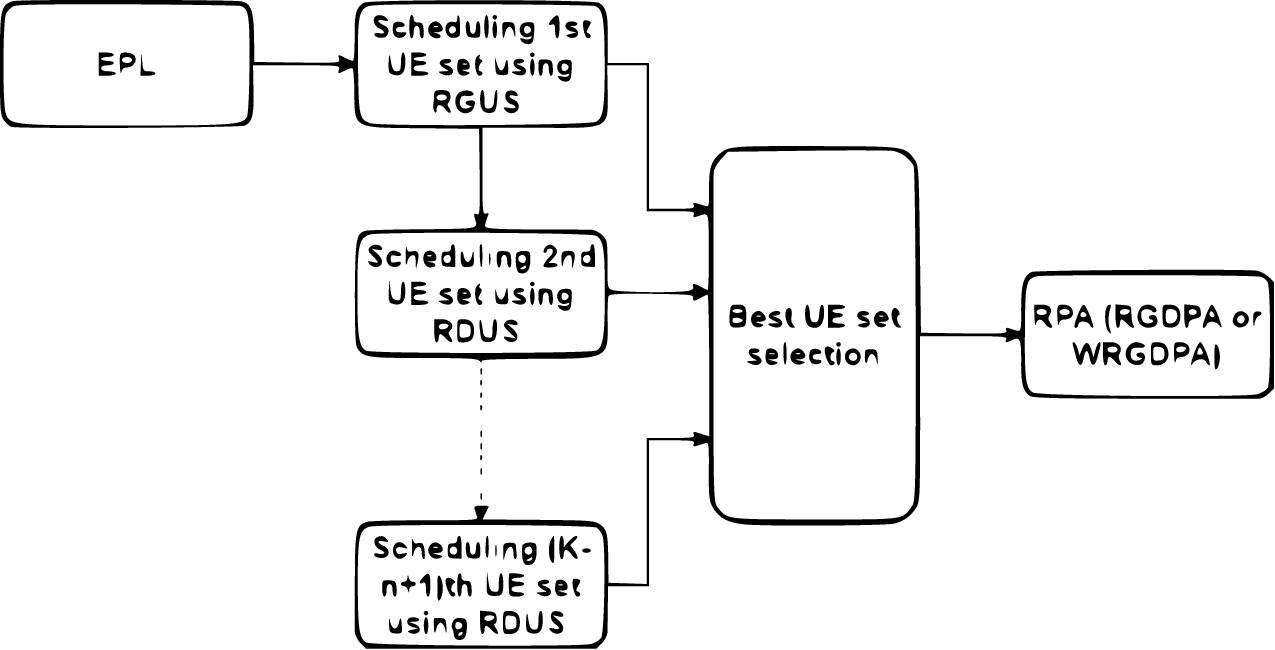}
	\caption{\small{Block diagram of the proposed robust resource allocation.}}
	\label{fig:Block Diagram}
\end{figure}

{In this section, we present the proposed robust resource allocation framework that is depicted in Fig. \ref{fig:Block Diagram}, which employs robust multiuser scheduling and power allocation algorithms operating sequentially. 
 
The proposed framework employs a robust greedy user scheduling (RGUS) technique to initially select a set of UEs under the assumption of equal power loading (EPL). The adoption of a greedy technique is due to its low computational cost and reliable performance, which makes it an effective and practical solution for optimization.  RGUS uses a robust dynamic UE substitution (RDUS) technique to consider potential UE sets and approach the optimal set identified}. Following the selection of the best UE set based on the sum-rate criterion, our framework incorporates a robust power allocation (RPA) strategy, which is designed to allocate the power efficiently, further enhancing the performance \footnote{This work focuses on robust resource allocation with standard MMSE and ZF precoders, which are not optimized for imperfect CSI at the transmitter. Exploring robust precoding techniques could further enhance performance and is a potential topic for future research.}.

\subsection{Robust multiuser scheduling}

The proposed robust multiuser scheduling algorithm based on worst-case optimization aims to solve the problem:
\begin{equation} \label{R.O.P}
\begin{aligned}
& \underset{\mathcal{S}_{n}}{\text{max}}~\underset{\tilde{\mathbf{G}}_a }{\text{min}}~SR\left ( \mathcal{S}_{n} \right ) \\
& \text{subject to} \left \| \mathbf{P}_a\left ( \mathcal{S}_{n} \right ) \right \|_{F}^{2}\leq P,\\
& \quad \ 
\beta\leq \left\| \tilde{\mathbf{g}}_{k}\right\|^{2}\leq \beta_0 , \forall k \in \mathcal{S}_{n}.
\end{aligned}
\end{equation}
where $\tilde{\mathbf{G}}_a$ is the channel estimation error, $ \tilde{\mathbf{g}}_{k}$ is the channel estimation error for UE \(k\), and $\beta$ and $\beta_0$ bound the estimation error vector power for each UE in the selected set. The problem in (\ref{R.O.P}) aims to optimize the sum-rate while targeting worst-case scenarios. {It introduces a minimization operation over $\tilde{\mathbf{G}}_a$, which represents the uncertainty on CSI. In contrast to \eqref{rop1}, this dual-layered optimization ensures that the sum-rate is maximized while the robustness against the worst possible channel outcomes is enforced. {The robust design assumes that $\tilde{\mathbf{G}}_a$ has components $\tilde{\mathbf{g}}_k$ that can take any value within the uncertainty region defined by $\beta\leq \left\| \tilde{\mathbf{g}}_{k}\right\|^{2}\leq \beta_0$ for all $k \in \mathcal{S}_n$, ensuring system performance does not rely on specific statistical outcomes but remains robust for all allowable error realizations.} As it will be shown later, using $\tilde{\mathbf{G}}_a$ leads to the optimization of the parameter $\alpha$ that models the uncertainty on CSI.}

In order to solve the problem in (\ref{R.O.P}), inspired by the clustered enhanced subset greedy (C-ESG) multiuser scheduling algorithm in \cite{mashdour2022enhanced}, we propose the robust C-ESG (RC-ESG) multiuser scheduling algorithm. We start by using the RGUS technique to schedule the first UE set. It first identifies a UE with the highest channel power by solving the robust extremum optimization problem (REOP) described by
\begin{equation} \label{rop2}
\begin{aligned}
k_{1}=& \underset{k\in \mathcal{K}}{\textup{argmax}} ~\left ( \underset{\tilde{\mathbf{g}}_{k}}{\min} \left ( \mathbf{g}_{{k}}^{H}\mathbf{g}_{{k}} \right )\right )\\ \
& \text{subject to}~ \beta\leq \left\| \tilde{\mathbf{g}}_{k}\right\|^{2}\leq \beta_0 , \forall k\in \mathcal{K},
\end{aligned}
\end{equation}
where $\mathcal{K}=\left \{ 1, \cdots , K \right \}$ is set of all UEs. The robust design of this problem ensures that the solution is robust to variations within the channel constraints. {Then, we select UEs to enhance the sum-rate. In each iteration, we identify and incorporate a UE that contributes to maximizing the channel power by solving a REOP problem similar to (\ref{rop2}) as given by}
\begin{equation}
\begin{aligned}
{k_{1}=}& {\arg \max_{k \in (\mathcal{K} \setminus \mathcal{U})} \left( \min_{\tilde{\mathbf{g}}_{k}} \left ( \mathbf{g}_{{k}}^{H}\mathbf{g}_{{k}} \right ) \right)}\\ \
& {\text{subject to}} \ {\beta \leq \left\| \tilde{\mathbf{g}}_{k} \right\|^2 \leq \beta_0, \forall k\in \mathcal{K} \setminus \mathcal{U},}
\end{aligned}
\end{equation}
{where \(\mathcal{U}\) denotes the set of UEs selected in previous iterations.}

This process repeats until we find $n$ UEs or it terminates when adding a new UE no longer improves the sum-rate. The set of selected UEs is represented by ${S}_{n}$. {Next, in an iterative manner, we use an RDUS technique to explore more UE sets}. In this regard, in each iteration $j$, we identify a UE denoted as $k_{r}(j)$ within the current set ${S}_{n}(j)$ that under the worst-case channel conditions, exhibits the least channel power.  Consequently, it is identified as the weakest link in terms of contribution to the sum-rate. This is achieved through the solution of the robust optimization problem expressed by 
\begin{equation} \label{rop3}
\begin{aligned}
k_{r}(j)=& \underset{k\in \mathcal{S}_{n}(j)}{\textup{argmin}} ~\left ( \underset{\tilde{\mathbf{g}}_{k}}{\max} \left ( \mathbf{g}_{{k}}^{H}\mathbf{g}_{{k}} \right )\right )\\ \
& \text{subject to} \ \beta\leq \left\| \tilde{\mathbf{g}}_{k}\right\|^{2}\leq \beta_0, \forall k\in \mathcal{S}_{n}(j).
\end{aligned}
\end{equation}  
The above optimization ensures the identification of the UE that could adversely affect the overall system performance under the worst estimation error. Next, we select a UE from the current set of the remaining UEs $\mathcal{K}_{re}(j)$, those UEs not yet chosen in previous iterations, in a manner that is robust against imperfect CSI via the optimization problem given by
\begin{equation} \label{rop4}
\begin{aligned}
k_{su}(j)=& \underset{k\in \mathcal{K}_{re}(j)}{\textup{argmax}} ~\left ( \underset{\tilde{\mathbf{g}}_{k}}{\min} \left ( \mathbf{g}_{{k}}^{H}\mathbf{g}_{{k}} \right )\right )\\ \
& \text{subject to} \ \beta\leq \left\| \tilde{\mathbf{g}}_{k}\right\|^{2}\leq \beta_0 , \forall k\in \mathcal{K}_{re}(j).
\end{aligned}
\end{equation} 
Note that $\mathcal{K}_{re}(1)=\mathcal{K}\setminus{S}_{n}(1)$. Then, we replace $k_{r}(j)$ in the current UE set by $k_{su}(j)$, obtaining the new UE set. 
Then, we update the current set of selected UEs and the current set of remaining UEs in each iteration as follows, respectively,
\begin{equation}
    \mathcal{S}_{n}(j)=\left (\mathcal{S}_{n}(j-1)\setminus k_{r}(j-1) \right )\cup k_{su}(j-1),
\end{equation}
\begin{equation}
    \mathcal{K}_{\text{re}}(j)=\mathcal{K}_{\text{re}}(j-1)\setminus k_{su}(j-1).
\end{equation}
This iteration is repeated and continues until we achieve a sufficient number of UE sets, allowing us to approximate the optimal set obtained through exhaustive search with reduced computational cost. The proposed RC-ESG algorithm is detailed in Algorithm \ref{alg:robust_alg}. {This algorithm's selection criterion ensures robust, near-optimal convergence, with dynamic error bounds adapting to changing conditions. Its stopping criterion prevents overfitting, leading to a stable local optimum with reduced complexity, as validated by numerical results.}

\SetKwInput{KwInput}{Input}
\SetKwInput{KwOutput}{Output}
\SetKwInput{KwInitialization}{Initialization}
\SetKw{KwGoTo}{go to}
\SetKwComment{Comment}{/* }{ */}

\begin{algorithm}
\caption{RC-ESG Scheduling Algorithm 
}\label{alg:robust_alg}

\KwInput{$\mathcal{K}, \beta, \beta_0, n, \text{system parameters}$}
\KwOutput{$\mathcal{S}_{n}^{d}$}

\KwInitialization{}
\Begin{
    Initialize $j=1$, $l = 1$\;
    Adjust $\beta, \beta_0$ based on initial system state\;
    Find a user such that 
    $k_{1}=\arg \max_{k\in \mathcal{K}} \left (  \min_{\tilde{\mathbf{g}}_{k}} \left ( \mathbf{g}_{{k}}^{H}\mathbf{g}_{{k}} \right )  \right )$ , \quad 
    $\text{s.t. } \beta\leq \left\| \tilde{\mathbf{g}}_{k}\right\|^{2}\leq \beta_0 , \forall k \in \mathcal{K}$
    and denote $k_{1}=\mathcal{U}_{1}$ and the rate $SR(\mathcal{U}_{1})$\;
}
\If{$l < n$}{
    $l = l+1$\;
    Adjust $\beta, \beta_0$ based on system performance\;
    Find a user $k_{l}$ such that
    
    {$k_{l} = \arg \max_{k \in (\mathcal{K} \setminus \mathcal{U}_{l-1})} \left( \min_{\tilde{\mathbf{g}}_{k}} \left ( \mathbf{g}_{{k}}^{H}\mathbf{g}_{{k}} \right ) \right)$, \quad 
    \text{s.t. } $\beta \leq \left\| \tilde{\mathbf{g}}_{k} \right\|^2 \leq \beta_0$}\;
    Set $\mathcal{U}_{l}=\mathcal{U}_{l-1} \cup {k_{l}}$ and denote the rate $SR(\mathcal{U}_{l})$\;
    \If{$SR(\mathcal{U}_{l}) \leq SR(\mathcal{U}_{l-1})$}{
        \KwGoTo end\;
    }

}

\For{$j=1$ \KwTo $K-n+1$}{
 \uIf{$j=1$}{
$\mathcal{S}_{n}(j)=\mathcal{U}_{l}$\;
$\mathcal{K}_{\text{re}}(j)=\mathcal{K}\setminus \mathcal{S}_{n}(j)$\;

 }
 \Else{
        $\mathcal{S}_{n}(j)=\left (\mathcal{S}_{n}(j-1)\setminus k_{r}(j-1) \right )\cup k_{su}(j-1)$\;
        $\mathcal{K}_{\text{re}}(j)=\mathcal{K}_{\text{re}}(j-1)\setminus k_{su}(j-1)$\;

    }
    Update $\beta, \beta_0$ based on current system state;
    $k_{r}(j)=\underset{k\in \mathcal{S}_{n}(j)}{\textup{argmin}}\left ( \underset{\tilde{\mathbf{g}}_{k}}{\max} \left ( \mathbf{g}_{{k}}^{H}\mathbf{g}_{{k}} \right )\right )$,  \quad 
    $\text{s.t. } \beta\leq \left\| \tilde{\mathbf{g}}_{k}\right\|^{2}\leq \beta_0 , \forall k\in \mathcal{S}_{n}(j)$;
    
 $k_{su}(j)=\underset{k\in \mathcal{K}_{\text{re}}(j)}{\textup{argmax}}~\left ( \underset{\tilde{\mathbf{g}}_{k}}{\min} \left ( \mathbf{g}_{{k}}^{H}\mathbf{g}_{{k}} \right )\right )$,  \quad 
    $\text{s.t. }\beta\leq \left\| \tilde{\mathbf{g}}_{k}\right\|^{2}\leq \beta_0 , \forall k\in \mathcal{K}_{\text{re}}(j)$;
    
 Compute: $SR(\mathcal{S}_{n}(j))$\;

}

$\mathcal{S}_{n}^{d}=\arg \max_{\mathcal{S}_{n} \in \mathcal{S}_{n}(m)} \left \{ SR\left ( \mathcal{S}_{n} \right ) \right \}$\;
\end{algorithm}

In Algorithm \ref{alg:robust_alg}, the focus is on the REOP problems similar to those in (\ref{rop2})-(\ref{rop4}). The analysis of the proposed solutions to these problems are discussed in Section \ref{Analysis}.

\subsection{Robust Power Allocation} \label{RPA1}

{Due to the computational cost and the difficulty to apply worst-case optimization for power allocation with the MMSE precoder, as discussed in Section \ref{WRGDPA}, we devise an alternative approach based on the contribution of the channel estimation matrix.} Next, we adopt the worst-case optimization strategy for a simpler precoder, i.e., ZF, which in this case offers a simpler analysis. Accordingly, we propose RGDPA and WRGDPA robust power allocation algorithms against the CSI errors which are used to determine the power allocation matrix $ \mathbf{D} $. The proposed algorithms are based on gradient descent techniques, which are robust and cost-effective. 

\subsubsection{RGDPA approach} \label{RPA1s}
 
Under imperfect CSI, the RGDPA algorithm incorporates the channel estimation error and considers the robust optimization given by
 \begin{equation} \label{R.PA.c}
\begin{aligned}
& {\underset{\mathbf{d}}{\text{min}}~\mathbb{E}\left [ \varepsilon \mid \hat{\mathbf{G}}_a \right ]} \\
& {\text{subject to}  \left\| \mathbf{W}\text{diag}\left( \mathbf{d} \right) \right\|^{2}\leq P.} 
\end{aligned}
\end{equation}
This approach accounts for the uncertainty in channel estimation, because by conditioning the expectation of the error $\varepsilon$ on the estimated channel matrix $\hat{\mathbf{G}}_a$, the optimization problem takes into account the presence of estimation errors leading to a power allocation that is less sensitive to imperfect CSI. 

In this context, we examine the error equation obtained in Appendix \ref{Error Expression} as (\ref{eq:trerr}), where $\mathbf{x} \sim \mathcal{CN}(\mathbf{0}, \mathbf{I}_{n})$ and $\mathbf{w} \sim \mathcal{CN}(0, \sigma_{w}^{2} \mathbf{I}_{n})$. We consider the elements of the error $\tilde{\mathbf{G}}_a$ as zero-mean random variables. Moreover, these elements are assumed to be mutually independent. Under these conditions, the expectation of the error given $\hat{\mathbf{G}}_a$ can be simplified as
\begin{equation} \label{exp.err.con}
  \begin{split}
  \mathbb{E}\left [ \varepsilon \mid \hat{\mathbf{G}}_a \right ]=
    &n+n\sigma _{w}^{2}+\\
    &{Tr}\left (\rho _{f}\textup{diag}\left ( \mathbf{d}\right )\mathbf{W}^{H}\hat{\mathbf{G}}_{a}^*\hat{\mathbf{G}}_{a}^T\mathbf{W}\textup{diag}\left ( \mathbf{d}\right ) \right )+\\&{{Tr}\left (\rho_{f}\mathbb{E}_{\tilde{\mathbf{G}}_a}\left[\mathbf{P}_{a}^{H}\tilde{\mathbf{G}}_{a}^*\tilde{\mathbf{G}}_{a}^T\mathbf{P}_{a}\right]\right )}-\\
    &{Tr}\left (\sqrt{\rho _{f}}\hat{\mathbf{G}}_{a}^T\mathbf{W}\textup{diag}\left ( \mathbf{d}\right ) \right )-\\
    &{Tr}\left (\sqrt{\rho _{f}}\textup{diag}\left ( \mathbf{d}\right )\mathbf{W}^{H}\hat{\mathbf{G}}_{a}^* \right ).
    \end{split}
  \end{equation}
{This simplification is achieved by expanding the error 
and applying trace operations. By considering \(\mathbf{x}\) and \(\mathbf{w}\) as Gaussian variables, we derive the expected error given \(\hat{\mathbf{G}}_a\), accounting for estimation errors.} Accordingly, we first compute the derivative of the conditional error expectation with respect to the power allocation matrix. By employing $\frac{\partial \text{Tr}(\mathbf{AB})}{\partial \mathbf{A}} = \mathbf{B} \odot \mathbf{I}$, where $\mathbf{A}$ is a diagonal matrix and $\odot$ is the Hadamard product \cite{petersen2008matrix}, and the cyclic property of the trace operator, the first derivative of the objective function with respect to the power allocation factors is given by
\begin{equation} \label{first.or}
 \begin{split}
     \frac{\partial \mathbb{E}\left [ \varepsilon \mid \hat{\mathbf{G}}_a \right ]  }{\partial \mathbf{d}}=&2\rho _{f}\left ( \mathbf{W}^{H}\hat{\mathbf{G}}_{a}^*\hat{\mathbf{G}}_{a}^T\mathbf{W}\textup{diag}\left ( \mathbf{d}\right ) \right )\odot \mathbf{I}-\\
     &2\sqrt{\rho _{f}}Re\left \{ \mathbf{W}^{H}\hat{\mathbf{G}}_{a}^*\odot \mathbf{I} \right \}.
     \end{split}
 \end{equation}
We consider the conditional error expectation convex with respect to the power allocation matrix, as it will be shown in Section \ref{Analys.Er}.  Hence, we can use a gradient descent technique to solve the conditional MSE minimization with respect to the power allocation factors. Thus, we update the power allocation coefficients as follows:
 \begin{equation}
     \mathbf{d}\left ( i \right )=\mathbf{d}\left ( i-1 \right )-\lambda \frac{\partial \mathbb{E}\left [ \varepsilon \mid \hat{\mathbf{G}}_a \right ] }{\partial \mathbf{d}}\bigg|_{\mathbf{d} = \mathbf{d} \left ( i -1\right )},
\end{equation}
where $i$ is the iteration index, and $\lambda$ is the positive step size and $\frac{\partial \mathbb{E}\left [ \varepsilon \mid \hat{\mathbf{G}}_a \right ] }{\partial \mathbf{d}}$ is obtained as in (\ref{first.or}). In order for the transmit power constraint to be satisfied as $\left \| \mathbf{W} \textup{diag}\left ( \mathbf{d} \right ) \right \|^{2}=\left \| \mathbf{P}_{a} \right \|^{2}\leq P$, after the iterations, 
we scale the power allocation coefficients as using the power scaling factor given by 
 \begin{equation}
 \eta =\sqrt{\frac{{Tr}\left ( \mathbf{P}_{a}\mathbf{P}_{a}^{H} \right )}{{Tr}\left ( \mathbf{W}\textup{diag}\left ( \mathbf{d}.\mathbf{d} \right )\mathbf{W}^{H} \right )}}
\end{equation}
 A summary of the proposed RGDPA algorithm is presented in Algorithm \ref{alg:RPA}. Note that when the error expectation is not conditioned, it gives rise to the gradient descent power allocation (GDPA) algorithm. In this case, we use the error equation in (\ref{eq:trerr}), $\mathbf{x} \sim \mathcal{CN}(\mathbf{0}, \mathbf{I}_{n})$, $\mathbf{w} \sim \mathcal{CN}(0, \sigma_{w}^{2} \mathbf{I}_{n})$. 
 and both $\hat{\mathbf{G}}_a$ and $\tilde{\mathbf{G}}_a$ are treated as zero-mean random variables. Furthermore, all these elements are considered to be mutually independent. Thus, the error expectation is expressed by
\begin{equation} \label{exp.err}
  \begin{split}
  \mathbb{E}\left [ \varepsilon \right ]=
    &n+n\sigma _{w}^{2}+ {{Tr}\left (\rho_{f}\mathbb{E}_{\tilde{\mathbf{G}}_a}\left[\mathbf{P}_{a}^{H}\tilde{\mathbf{G}}_{a}^*\tilde{\mathbf{G}}_{a}^T\mathbf{P}_{a}\right]\right )}+\\
    &{Tr}\left (\rho _{f}\textup{diag}\left ( \mathbf{d}\right )\mathbf{W}^{H}\hat{\mathbf{G}}_{a}^*\hat{\mathbf{G}}_{a}^T\mathbf{W}\textup{diag}\left ( \mathbf{d}\right ) \right )
    \end{split}
  \end{equation}
 Hence, we can obtain the first derivative of the error expectation with respect to the power allocation factors as follows:
\begin{equation} \label{der1.err.exp} 
 \begin{split}
     \frac{\partial \mathbb{E}\left [ \varepsilon \right ]  }{\partial \mathbf{d}}=&2\rho _{f}\left ( \mathbf{W}^{H}\hat{\mathbf{G}}_{a}^*\hat{\mathbf{G}}_{a}^T\mathbf{W}\textup{diag}\left ( \mathbf{d}\right ) \right )\odot \mathbf{I}
     \end{split}
 \end{equation}
Accordingly, the GDPA algorithm is a special case of Algorithm \ref{alg:RPA}, where the power allocation factors are updated in each iteration using $\frac{\partial \mathbb{E}\left [ \varepsilon \right ]  }{\partial \mathbf{d}}$ as obtained in (\ref{der1.err.exp}). {The convexity of the conditional error expectation ensures that the GDPA algorithm converges to a global minimum, while the normalization step enforces the power constraint.}

\begin{algorithm}
\caption{RGDPA Power Allocation Algorithm}\label{alg:RPA}

\KwInput{$\mathbf{G}_{a}$, $\mathbf{P}_{a}$, $\mathbf{W}$, $\lambda$, $\mathbf{d}\left ( 1 \right )$, $\textup{I}_{D}$ }
 
    \For{$i = 2$ \KwTo $\textup{I}_{D}$}{
        $\frac{\partial \mathbb{E}\left [ \varepsilon \mid \hat{\mathbf{G}}_a \right ]  }{\partial \mathbf{d}}= 2\rho _{f}\left ( \mathbf{W}^{H}\hat{\mathbf{G}}_{a}^*\hat{\mathbf{G}}_{a}^T\mathbf{W}\textup{diag}\left ( \mathbf{d}\right ) \right )\odot \mathbf{I}- 2\sqrt{\rho _{f}}Re\left \{ \mathbf{W}^{H}\hat{\mathbf{G}}_{a}^*\odot \mathbf{I} \right \}$\;
        $\mathbf{d}\left ( i \right ) = \mathbf{d}\left ( i-1 \right )-\lambda \frac{\partial \mathbb{E}\left [ \varepsilon \mid \hat{\mathbf{G}}_a \right ]  }{\partial \mathbf{d}}\bigg|_{\mathbf{d} = \mathbf{d} \left ( i -1\right )}$\;
                }
\end{algorithm}

\subsubsection{WRGDPA approach} \label{WRGDPA}

In order to make the optimization problem given in (\ref{optmse.1}) robust against imperfect CSI, a promising alternative is to define the worst-case robust optimization problem as follows:
\begin{equation} \label{R.PA}
\begin{aligned}
& \hspace{6em} \underset{\mathbf{d}}{\text{min}}~\underset{\tilde{\mathbf{G}}_{a} }{\text{max}}~\mathbb{E}\left [ \varepsilon \right ] \\
& \text{subject to}  \left\| \mathbf{W} \text{diag}\left( \mathbf{d} \right) \right\|^{2}\leq P ~{\rm and}~  \beta_1\leq \left\| \tilde{\mathbf{G}}_{a}\right\|^{2}\leq \beta_2.
\end{aligned}
\end{equation}
In (\ref{R.PA}) the error expectation $\mathbb{E}\left [ \varepsilon \right ]$ in (\ref{exp.err}) is used. 

{Based on the analysis in Section \ref{Analys.Er.Ga} and the objective function in (\ref{R.PA}), in order to obtain robustness against imperfect CSI, the CSI imperfection level parameter $\alpha$ should be optimized.} Hence, for the convexity analysis of the objective function with respect to CSI imperfection level $\alpha$, we consider
\begin{equation} \label{GaVtild}
    \tilde{\mathbf{G}}_{a}=\sqrt{\alpha }\tilde{\mathbf{V}} , \tilde{\mathbf{V}} _{mk}=\sqrt{\beta_{mk}}\tilde{h}_{mk}
\end{equation}
\begin{equation}\label{GaV}
    \hat{\mathbf{G}}_{a}=\sqrt{\left ( 1-\alpha  \right )} \mathbf{V}, \mathbf{V} _{mk}=\sqrt{\beta_{mk}}h_{mk}
\end{equation}
Thus, we rewrite (\ref{exp.err}) as
\begin{equation} \label{exp.err.alpha}
  \begin{split}
  \mathbb{E}\left [ \varepsilon \right ]=
    &n+n\sigma _{w}^{2}+\\
    &{Tr}\left (\rho _{f}\left ( 1-\alpha  \right )\textup{diag}\left ( \mathbf{d}\right )\mathbf{W}^{H}{\mathbf{V}}^*{\mathbf{V}}^T\mathbf{W}\textup{diag}\left ( \mathbf{d}\right ) \right )+\\
    &{{Tr}\left (\rho_{f}\alpha\mathbb{E}_{\tilde{\mathbf{V}}}\left[\mathbf{P}_{a}^{H}\tilde{\mathbf{V}}^*\tilde{\mathbf{V}}^T\mathbf{P}_{a}\right]\right )}
    \end{split}
  \end{equation}
We consider $\mathbf{P}_{a}$ as the linear MMSE precoder given by
\begin{equation}
    \mathbf{P}_{a}=\eta \mathbf{L} ^{-1}\mathbf{G}_{a}^{H}\mathbf{F}^{H}
\end{equation}
where 
\begin{equation}
    \eta=\sqrt{\frac{P}{Tr\left ( \mathbf{L}^{-2} \mathbf{G}_{a}^{H}\mathbf{F}^{H}\mathbf{C}_{\mathbf{x}}\mathbf{F}\mathbf{G}_{a}\right )}},
\end{equation}
\begin{equation}
    \mathbf{L}=\mathbf{G}_{a}^{H}\mathbf{F}^{H}\mathbf{F}\mathbf{G}_{a}+\frac{Tr\left ( \mathbf{F}\mathbf{C}_{\mathbf{w}}\mathbf{F}^{H} \right )}{P}\mathbf{I}_{n},
\end{equation}
and $\mathbf{F}$ is a linear equalizer matrix which we consider as $\mathbf{F}=\mathbf{I}_{n}$, $\mathbf{I}_{n}\in \mathbb{C}^{n\times n}$ is the identity matrix, $\mathbf{C}_{\mathbf{x}}=\mathbf{I}_{n}$ is the signal covariance matrix, and $\mathbf{C}_{\mathbf{w}}=\sigma_{w}^{2}\mathbf{I}_{n}$ is the noise covariance matrix. The application of the MMSE linear precoder leads to intricate expressions using the derivative rules, posing challenges for mathematical tractability regarding the first and the second derivatives of (\ref{exp.err.alpha}) with respect to $\alpha$. Therefore, in order to be able to perform the worst-case optimization analysis, we employ a linear ZF precoder.

In this regard, the ZF precoder is described by
\begin{equation}
    \mathbf{P}_{a} = (\mathbf{G}_{a}^H \mathbf{G}_{a})^{-1} \mathbf{G}_{a}^H
\end{equation}
Using Equations (\ref{GaVtild}) and (\ref{GaV}), we can rewrite $\mathbf{P}_{a} $ as
\begin{equation}
    \mathbf{P}_{a} =\left ( \mathbf{H} \right )^{-1}\mathbf{Q}, 
\end{equation}
where
\begin{equation}
    \begin{split}
        \mathbf{H}=&(1 - \alpha) \mathbf{V}^H \mathbf{V} + \sqrt{\alpha(1 - \alpha)} \mathbf{V}^H \tilde{\mathbf{V}} \\
       & + \sqrt{\alpha(1 - \alpha)} \tilde{\mathbf{V}}^H \mathbf{V} + \alpha \tilde{\mathbf{V}}^H \tilde{\mathbf{V}},
    \end{split}
\end{equation}
\begin{equation}
    {\mathbf{Q}= \sqrt{\alpha}\tilde{\mathbf{V}}^H+\sqrt{(1 - \alpha)}\mathbf{V}^H.} 
\end{equation}
 Calculating the first derivatives of $\mathbf{H}$ and $\mathbf{Q}$ with respect to $\alpha$, we obtain the following equations:
 \begin{equation}
 \begin{split}
     \mathbf{H}_{d1}=\frac{d\mathbf{H}}{d\alpha }=&-\mathbf{V}^H \mathbf{V}+\frac{1-2\alpha }{2\sqrt{\alpha \left ( 1-\alpha  \right )}}\mathbf{V}^H \tilde{\mathbf{V}}\\
     &+\frac{1-2\alpha }{2\sqrt{\alpha \left ( 1-\alpha  \right )}}\tilde{\mathbf{V}}^H \mathbf{V}+\tilde{\mathbf{V}}^H \tilde{\mathbf{V}},
     \end{split}
 \end{equation}
 \begin{equation}
     {\mathbf{Q}_{d1}=\frac{d \mathbf{Q}}{d \alpha} = \frac{1}{2\sqrt{\alpha}}\tilde{\mathbf{V}}^H - \frac{1}{2\sqrt{(1 - \alpha)}}\mathbf{V}^H.}
 \end{equation}
 To obtain the first derivative of the objective function $\mathbb{E}\left [ \varepsilon \right ]$ given in (\ref{exp.err.alpha}) with respect to $\alpha$, we use the chain rule for derivatives and $\frac{\partial \mathbf{A}^{-1}}{\partial x}=-\mathbf{A}^{-1}\frac{\partial \mathbf{A}}{\partial x}\mathbf{A}^{-1}$ where $\mathbf{A}$ is a matrix and $x$ is a scalar \cite{petersen2008matrix}, and we take the derivative of $\mathbf{P}_{a}$ as
 \begin{equation}
    {\mathbf{P}_{ad1}=\frac{\partial \mathbf{P}_{a}}{\partial \alpha }=-\mathbf{H}^{-1}\mathbf{H}_{d1}\mathbf{H}^{-1}\mathbf{Q}+\mathbf{H}^{-1}\mathbf{Q}_{d1}.}
 \end{equation}
 Consequently, we obtain the first derivative of the objective function with respect to $\alpha$ as follows:
 \begin{equation} \label{First,dev}
 \begin{split}
     {\frac{\partial \mathbb{E}\left [ \varepsilon \right ]}{\partial \alpha }=}&{{Tr}\left (\rho_{f}\mathbb{E}_{\tilde{\mathbf{V}}}\left[\mathbf{P}_{a}^{H}\tilde{\mathbf{V}}^*\tilde{\mathbf{V}}^T\mathbf{P}_{a}\right]\right )+}\\
     &{{Tr}\left (\rho_{f}\alpha\mathbb{E}_{\tilde{\mathbf{V}}}\left[\mathbf{P}_{ad1}^{H}\tilde{\mathbf{V}}^*\tilde{\mathbf{V}}^T\mathbf{P}_{a}\right]\right )+}\\
     &{{Tr}\left (\rho_{f}\alpha\mathbb{E}_{\tilde{\mathbf{V}}}\left[\mathbf{P}_{a}^{H}\tilde{\mathbf{V}}^*\tilde{\mathbf{V}}^T\mathbf{P}_{ad1}\right]\right )-}\\
     &{{Tr}\left (\rho _{f}\textup{diag}\left ( \mathbf{d}\right )\mathbf{W}^{H}\mathbf{V}^*\mathbf{V}^T\mathbf{W}\textup{diag}\left ( \mathbf{d}\right ) \right )}
     \end{split}
 \end{equation}
 For the second derivative of the objective function with respect to $\alpha$, we obtain derivatives of $\mathbf{H}_{d1}$, $\mathbf{Q}_{d1}$ and $\mathbf{P}_{ad1}$ as follows:
 \begin{equation}
     \mathbf{H}_{d2}=\frac{\partial \mathbf{H}_{d1}}{\partial \alpha }=\frac{-1}{4\alpha \left ( 1-\alpha  \right )\sqrt{\alpha \left ( 1-\alpha  \right )}}\left ( \mathbf{V}^H \tilde{\mathbf{V}}+\tilde{\mathbf{V}}^H \mathbf{V} \right ),
 \end{equation}
 \begin{equation}
     {\mathbf{Q}_{d2}=\frac{\partial \mathbf{Q}_{d1}}{\partial \alpha }=-\frac{1}{4\sqrt{\alpha ^{3}}}\tilde{\mathbf{V}}^H -\frac{1}{4\sqrt{ \left ( 1-\alpha  \right )^{3}}}\mathbf{V}^H,}
 \end{equation}
 \begin{equation}
     \begin{split}
         \mathbf{P}_{ad2}=\frac{\partial \mathbf{P}_{ad1}}{\partial \alpha }=&-\mathbf{H}_{d1}\mathbf{H}_{d1}\mathbf{H}^{-1}\mathbf{Q}-\mathbf{H}\mathbf{H}_{d2}\mathbf{H}^{-1}\mathbf{Q}\\
         &+\mathbf{H}\mathbf{H}_{d1}\mathbf{H}\mathbf{H}_{d1}\mathbf{H}^{-1}\mathbf{H}^{-1}\mathbf{Q}\\
         &-\mathbf{H}\mathbf{H}_{d1}\mathbf{H}^{-1}\mathbf{Q}_{d1}-\mathbf{H}\mathbf{H}_{d1}\mathbf{H}^{-1}\mathbf{Q}_{d1}\\
         &+\mathbf{H}^{-1}\mathbf{Q}_{d2}
     \end{split}
 \end{equation}
 Note that $\mathbf{H}_{d1}$ is a square matrix. Accordingly, the second derivative of $\mathbb{E}\left [ \varepsilon \right ]$ with respect to $\alpha$ could be written as
 \begin{equation} \label{Second,dev}
     \begin{split}
         {\frac{\partial^{2} \mathbb{E}\left [ \varepsilon \right ]}{\partial \alpha^{2} }=}&
         {{Tr}\left (\rho_{f}\mathbb{E}_{\tilde{\mathbf{V}}}\left[ \mathbf{P}_{ad1}^{H}\tilde{\mathbf{V}}^*\tilde{\mathbf{V}}^T\mathbf{P}_{a} \right]\right )+}\\&{{Tr}\left (\rho_{f}\mathbb{E}_{\tilde{\mathbf{V}}}\left[ \mathbf{P}_{a}^{H}\tilde{\mathbf{V}}^*\tilde{\mathbf{V}}^T\mathbf{P}_{ad1} \right]\right )+}\\
         &{{Tr}\left (\rho_{f}\mathbb{E}_{\tilde{\mathbf{V}}}\left[ \mathbf{P}_{ad1}^{H}\tilde{\mathbf{V}}^*\tilde{\mathbf{V}}^T\mathbf{P}_{a} \right]\right )+}\\&{{Tr}\left (\rho_{f}\alpha\mathbb{E}_{\tilde{\mathbf{V}}}\left[ \mathbf{P}_{ad2}^{H}\tilde{\mathbf{V}}^*\tilde{\mathbf{V}}^T\mathbf{P}_{a} \right]\right )+}\\
         &{{Tr}\left (\rho_{f}\alpha\mathbb{E}_{\tilde{\mathbf{V}}}\left[ \mathbf{P}_{ad1}^{H}\tilde{\mathbf{V}}^*\tilde{\mathbf{V}}^T\mathbf{P}_{ad1} \right]\right )+}\\&{{Tr}\left (\rho_{f}\mathbb{E}_{\tilde{\mathbf{V}}}\left[ \mathbf{P}_{a}^{H}\tilde{\mathbf{V}}^*\tilde{\mathbf{V}}^T\mathbf{P}_{ad1} \right]\right )+}\\
         &{{Tr}\left (\rho_{f}\alpha\mathbb{E}_{\tilde{\mathbf{V}}}\left[ \mathbf{P}_{ad1}^{H}\tilde{\mathbf{V}}^*\tilde{\mathbf{V}}^T\mathbf{P}_{ad1} \right]\right )+}\\&{{Tr}\left (\rho_{f}\alpha\mathbb{E}_{\tilde{\mathbf{V}}}\left[ \mathbf{P}_{a}^{H}\tilde{\mathbf{V}}^*\tilde{\mathbf{V}}^T\mathbf{P}_{ad2}\right]\right ).}
     \end{split}
 \end{equation}
 Now, in order to solve the maximization over channel estimation error in the optimization problem in (\ref{R.PA}), since we aim to solve the problem with respect to $\alpha$, we resort to: 
 \begin{itemize}
     \item $\frac{\partial^{2} \mathbb{E}\left [ \varepsilon  \right ]}{\partial \alpha^{2}}>0$: In this case, since $\mathbb{E}\left [ \varepsilon  \right ]$ is convex with respect to $\alpha$, any local extremum (minimum or maximum) within the feasible region is also a global extremum. For maximization problems with a convex objective, the global maximum occurs at the boundary of the feasible region. This concept is illustrated in Fig.~\ref{figcon}, where the convex function $f\left ( \textup{x} \right )$ is examined within the interval $\textup{x}_{1}\leq \textup{x}\leq \textup{x}_{2}$, revealing that its maximum value occurs precisely at the endpoint $\textup{x}_{2}$. 
    \begin{figure}
    \centering
\includegraphics[width=.725\linewidth]{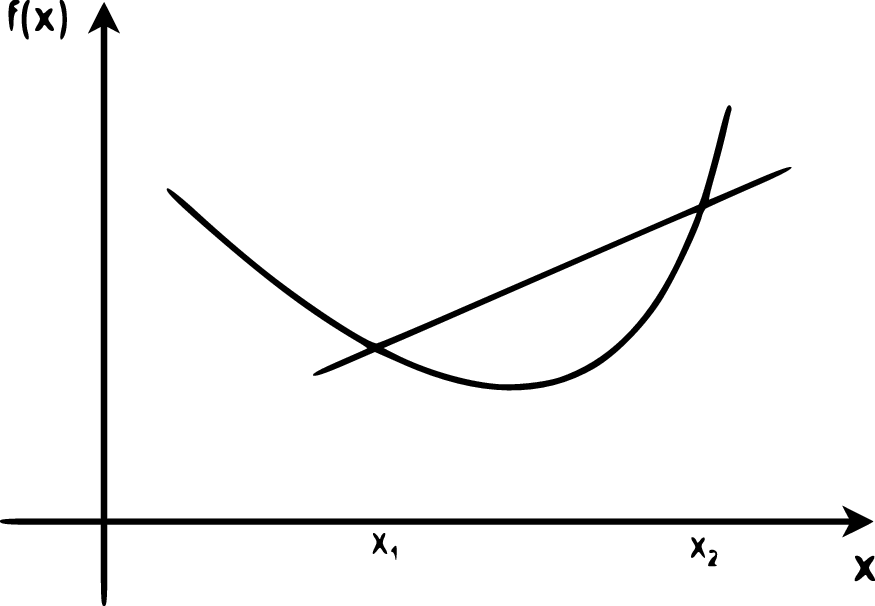}
        \vspace{-1.25em}
    \caption{Boundary extremum illustration, convex case}
    \label{figcon}
\end{figure} Thus, the maximum of the objective function will occur at the boundary of the feasible set and the desired $\alpha$ is given as follows:
\begin{align} \label{alph.d2>0}
        \alpha = 
        \begin{cases}
        \frac{\beta_1}{\| \tilde{\mathbf{V}} \|^{2}}, & \text{if } \ \mathbb{E}\left [ \varepsilon  \right ] \bigg|_{\alpha=\frac{\beta_1}{\| \tilde{\mathbf{V}} \|^{2}}} > \mathbb{E}\left [ \varepsilon  \right ] \bigg|_{\alpha=\frac{\beta_2}{\| \tilde{\mathbf{V}} \|^{2}}} \\
        \frac{\beta_2}{\| \tilde{\mathbf{V}} \|^{2}}, & \text{if }  \ \mathbb{E}\left [ \varepsilon  \right ]  \bigg|_{\alpha=\frac{\beta_2}{\| \tilde{\mathbf{V}} \|^{2}}} > \mathbb{E}\left [ \varepsilon  \right ] \bigg|_{\alpha=\frac{\beta_1}{\| \tilde{\mathbf{V}} \|^{2}}} 
        \end{cases}
    \end{align}

    \item $\frac{\partial^{2} \mathbb{E}\left [ \varepsilon  \right ]}{\partial \alpha^{2}}<0$: In this case, $\mathbb{E}\left [ \varepsilon  \right ]$ is concave  with respect to $\alpha$. Thus we use a gradient ascent technique to find the desired $\alpha$ which maximizes the objective $\mathbb{E}\left [ \varepsilon  \right ]$. In this regard, the imperfection parameter is updated as 
\begin{equation}
    \alpha \left ( t \right )=\alpha \left ( t -1\right )+\gamma \frac{\partial \mathbb{E}\left [ \varepsilon  \right ]}{\partial \alpha }\bigg|_{\alpha = \alpha \left ( t -1\right )}
\end{equation}
where $t$ is the iteration index, $\gamma$ is the positive step size and $\frac{\partial \mathbb{E}\left [ \varepsilon  \right ]}{\partial \alpha }$ is obtained using (\ref{First,dev}). After obtaining the desired channel estimation error parameter as $\alpha_d$, if the feasible set is not met, the following update is to ensure that it lies in the feasible set $\beta_1\leq \left\| \tilde{\mathbf{G}}_{a}\right\|^{2}\leq \beta_2$:
\begin{align} \label{projection.PA}
        \alpha_d= 
        \begin{cases}
        \frac{\beta_1}{\| \tilde{\mathbf{V}} \|^{2}}, & \text{if } \alpha < \frac{\beta_1}{\| \tilde{\mathbf{V}} \|^{2}} \\
        \frac{\beta_2}{\| \tilde{\mathbf{V}} \|^{2}}, & \text{if } \alpha > \frac{\beta_2}{\| \tilde{\mathbf{V}} \|^{2}}
        \end{cases}
    \end{align}
 \end{itemize}
Next, we solve the minimization of the MSE over the power loading matrix $\mathbf{D}$. We first investigate the convexity of $\mathbb{E}\left [ \varepsilon \right ]$ given in (\ref{exp.err}). Then, we derive the first derivative with respect to $\mathbf{d}$ as (\ref{der1.err.exp}). However, as (\ref{first.or}) and (\ref{der1.err.exp}) are the same with respect to the power allocation factors $\mathbf{d}$, the convexity analysis of the error expectation over $\mathbf{d}$ is the same as that in Section \ref{Analys.Er}. Thus, we conclude that the objective $\mathbb{E}\left [ \varepsilon \right ]$ is convex with respect to the power allocation factors. Therefore, in order to obtain the MSE minimization solution with respect to the power allocation factors, we use the same gradient descent technique developed in Section \ref{RPA1}, while $\frac{\partial \mathbb{E}\left [ \varepsilon \right ]  }{\partial \mathbf{d}}$ is calculated using (\ref{der1.err.exp}) to update the power allocation.

 We can now detail the proposed worst-case robust gradient descent power allocation (WRGDPA) algorithm in Algorithm \ref{alg:RPA2}, where $\textup{I}_{G}$ and $\textup{I}_{D}$ show the numbers of iterations in the iterative maximization and iterative minimization in the solution to the problem in (\ref{R.PA}), respectively. {The WRGDPA algorithm ensures optimality by identifying the worst-case $\alpha$ based on convexity and then minimizing the error over $\mathbf{d}$ using gradient descent, while enforcing power constraints. At the receiver,  algorithms for detection and interference mitigation
\cite{jidf,rrber,spa,mfsic,mbdf} can be employed.
}

 \begin{algorithm}
\caption{WRGDPA Power Allocation Algorithm}
\label{alg:RPA2}

\KwInput{$\mathbf{G}_{a}$, $\mathbf{P}_{a}$, $\mathbf{W}$, $\tilde{\mathbf{V}}$, $\mathbf{V}$, $\lambda$, $\gamma$, $\beta_1$, $\beta_2$, $\textup{I}_{D}$, $\textup{I}_{G}$, $\alpha\left ( 1 \right )$, $\mathbf{d}\left ( 1 \right )$}

Calculate $\frac{\partial \mathbb{E}[\varepsilon]}{\partial \alpha}$ using Eq. (\ref{First,dev}) and $\frac{\partial^{2} \mathbb{E}\left [ \varepsilon  \right ]}{\partial \alpha^{2}}$ as Eq. (\ref{Second,dev})\;

\eIf{$\frac{\partial^{2} \mathbb{E}\left [ \varepsilon  \right ]}{\partial \alpha^{2}}>0$}{
    Derive $\alpha$ as Eq. (\ref{alph.d2>0})\;
    $\tilde{\mathbf{G}}_{a}=\sqrt{\alpha}\tilde{\mathbf{V}}$\;
    $\hat{\mathbf{G}}_{a}=\sqrt{1-\alpha}\mathbf{V}$\;
}{
    \For{$t = 2$ \KwTo $\textup{I}_{G}$}{
        $\alpha \left ( t \right )=\alpha \left ( t -1\right )+\gamma \frac{\partial \mathbb{E}\left [ \varepsilon  \right ]}{\partial \alpha }\bigg|_{\alpha = \alpha \left ( t -1\right )}$\;
        
    }
    $\alpha_d=\alpha(\textup{I}_{G})$\;
    \uIf{$\alpha_d < \frac{\beta_1}{\| \tilde{\mathbf{V}} \|^{2}}$}{
            $\alpha_d=\frac{\beta_1}{\| \tilde{\mathbf{V}} \|^{2}}$\;
        }
        \uElseIf{$\alpha_d > \frac{\beta_2}{\| \tilde{\mathbf{V}} \|^{2}}$}{
            $\alpha_d=\frac{\beta_2}{\| \tilde{\mathbf{V}} \|^{2}}$\;
        }
        $\tilde{\mathbf{G}}_{a}=\sqrt{\alpha_d}\tilde{\mathbf{V}}$\;
        $\hat{\mathbf{G}}_{a}=\sqrt{1-\alpha_d}\mathbf{V}$\;
}

\For{$i = 2$ \KwTo $\textup{I}_{D}$}{
    $\frac{\partial \mathbb{E}\left ( \varepsilon  \right ) }{\partial \mathbf{D}}=2\rho _{f}\left ( \mathbf{W}^{H}\hat{\mathbf{G}}_{a}^*\hat{\mathbf{G}}_{a}^T\mathbf{W}\textup{diag}\left ( \mathbf{d}\left ( {i-1} \right )\right ) \right )\odot \mathbf{I}$\;
    $\mathbf{d}\left ( i \right )=\mathbf{d}\left ( i-1 \right )-\lambda \frac{\partial \mathbb{E}\left ( \varepsilon  \right ) }{\partial \mathbf{d}}\bigg|_{\mathbf{d} = \mathbf{d} \left ( i -1\right )}$\;
    
}

  $\mathbf{d}_d=\mathbf{d}\left ( \textup{I}_{D} \right )$ \;

\If{${Tr}\left ( \mathbf{W}\textup{diag}\left ( \mathbf{d}_d.\mathbf{d}_d \right )\mathbf{W}^{H} \right ) \neq {Tr}\left ( \mathbf{P}_{a}\mathbf{P}_{a}^{H} \right )$}{
        $\eta =\sqrt{\frac{{Tr}\left ( \mathbf{P}_{a}\mathbf{P}_{a}^{H} \right )}{{Tr}\left ( \mathbf{W}\textup{diag}\left ( \mathbf{d}_d.\mathbf{d}_d \right )\mathbf{W}^{H} \right )}}$\;
        $\mathbf{d}_d=\eta \mathbf{d}_d$\;
    }
\end{algorithm}

\section{Analysis} \label{Analysis}

This section presents an analysis of convexity and introduces solutions for the REOP problems ((\ref{rop2})-(\ref{rop4})) in Algorithm \ref{alg:robust_alg}. It then examines the convexity of the objective function of the robust power allocation problem in (\ref{R.PA.c}), and the convexity analysis of the error expectation with respect to the channel estimation error $\tilde{\mathbf{G}}_a$. An analysis of complexity of the proposed and existing approaches is also provided.

\subsection{Analysis and Proposed Solutions to REOP Problems}
\subsubsection{Convexity analysis of the objective function $J$} 

For user \( k \), the channel from AP \( m \) is denoted as \( g_{mk} \) as specified in (\ref{eq.gI}). Consequently, the channel vector for user \( k \) is defined by
\begin{equation}
   \mathbf{g}_{{k}} = \left[ g_{{1k}}, \cdots, g_{{Mk}} \right]^T,
\end{equation}
where \( \mathbf{g}_k \) contains the channel coefficients from all \( M \) APs to user \( k \). Using (\ref{eq.gI}), we rewrite $\mathbf{g}_{{k}}$ as follows:
\begin{equation} \label{chan.vec.k}
    \mathbf{g}_{{k}}=\hat{\mathbf{g}}_{k}+\tilde{\mathbf{g}}_{k},
\end{equation}
where $\hat{\mathbf{g}}_{k}$ is the channel estimation vector and $\tilde{\mathbf{g}}_{k}$ is the channel estimation error vector for UE $k$. 

Next, we rewrite the objective function \(J\) as:
\begin{equation} \label{objective}
\begin{split}
    J =& \mathbf{g}_{{k}}^{H}\mathbf{g}_{{k}} = \hat{\mathbf{g}}_{k}^{H}\hat{\mathbf{g}}_{k} + \hat{\mathbf{g}}_{k}^{H}\tilde{\mathbf{g}}_{k} + \tilde{\mathbf{g}}_{k}^{H}\hat{\mathbf{g}}_{k} + \tilde{\mathbf{g}}_{k}^{H}\tilde{\mathbf{g}}_{k} = \\
    & \hat{\mathbf{g}}_{k}^{H}\hat{\mathbf{g}}_{k} + 2\operatorname{Re}\left ( \hat{\mathbf{g}}_{k}^{H}\tilde{\mathbf{g}}_{k} \right ) + \tilde{\mathbf{g}}_{k}^{H}\tilde{\mathbf{g}}_{k}.
\end{split}
\end{equation}
Given that \(J\) is scalar, it can be equated to its trace. Consequently, to compute the first derivative of \(J\) with respect to \(\tilde{\mathbf{g}}_{k}\), we employ the following equation \cite{petersen2008matrix}:
\begin{equation} \label{der2}
    \frac{d f\left ( z \right )}{dz} = \frac{1}{2}\left ( \frac{\partial f\left ( z \right )}{\partial \operatorname{Re}\left ( z \right )} - j\frac{\partial f\left ( z \right )}{\partial \operatorname{Im}\left ( z \right )} \right ),
\end{equation}
where \(z\) is a complex variable, \(\operatorname{Re}\left ( . \right )\) and \(\operatorname{Im}\left ( . \right )\) are the real and imaginary parts, respectively, and \(j\) is the imaginary unit. The vector differentiation rule is then adopted \cite{barnes2006matrix}: 
\begin{equation} 
     \frac{\partial \mathbf{a}}{\partial \mathbf{b}} = \begin{bmatrix}
\frac{\partial a_{1}}{\partial b_{1}} & \cdots  & \frac{\partial a_{m}}{\partial b_{1}} \\ 
\vdots & \ddots & \vdots \\ 
\frac{\partial a_{1}}{\partial b_{n}} & \cdots & \frac{\partial a_{m}}{\partial b_{n}}
\end{bmatrix},
\end{equation}
where \(\mathbf{a}\) is an m-element vector, and \(\mathbf{b}\) is an n-element vector. Consequently, we deduce that
\begin{equation} \label{der3}
     \frac{\partial {a}}{\partial \mathbf{b}} = \begin{bmatrix}
\frac{\partial a}{\partial b_{1}} \\ 
\vdots \\ 
\frac{\partial a}{\partial b_{n}}
\end{bmatrix},
\end{equation}
where \(a\) is a scalar. As such, \(\frac{\partial \hat{\mathbf{g}}_{k}^{H}\hat{\mathbf{g}}_{k}}{\partial \tilde{\mathbf{g}}_{k}} = 0\), and
by applying (\ref{der2}) and (\ref{der3}), it can be demonstrated that 
\begin{equation} \label{rond.real}
\frac{\partial \left ( 2\operatorname{Re}\left ( \hat{\mathbf{g}}_{k}^{H}\tilde{\mathbf{g}}_{k} \right )
 \right )
}{\partial \tilde{\mathbf{g}}_{k}} = \hat{\mathbf{g}}_{k}^{H}.
\end{equation}

Moreover, by invoking \(\frac{  \partial\mathbf{x}^{H}\mathbf{x} }{\partial \mathbf{x}} = \mathbf{x}^{H}\), as per \cite{haykin2002adaptive}, we obtain
\begin{equation}
    \frac{\partial \left ( \tilde{\mathbf{g}}_{k}^{H}\tilde{\mathbf{g}}_{k} \right )}{\partial \tilde{\mathbf{g}}_{k}} = \tilde{\mathbf{g}}_{k}^{H}.
\end{equation}
Therefore, the first derivative of the objective function in relation to \(\tilde{\mathbf{g}}_{k}\) is 
\begin{equation} \label{der.first}
    \frac{\partial J}{\partial \tilde{\mathbf{g}}_{k}} = \hat{\mathbf{g}}_{k}^{H} + \tilde{\mathbf{g}}_{k}^{H}.
\end{equation}
The second derivative with respect to \(\tilde{\mathbf{g}}_{k}\) is given by 
\begin{equation} \label{der.sec}
    \frac{\partial^{2} J}{\partial \tilde{\mathbf{g}}_{k}^{2}} = \mathbf{0},
\end{equation}
indicating that the objective function is affine with respect to \(\tilde{\mathbf{g}}_{k}\). This holds even when the derivative is taken with respect to \(\hat{\mathbf{g}}_{k}\). Furthermore, when the derivative is calculated with respect to the small scale fading error vector \(\tilde{\mathbf{h}}_{k} = [ \tilde{h}_{{1k}}, \cdots, \tilde{h}_{{Mk}} ]^T\), the second derivative is also zero, as explained in Appendix \ref{der2toh}. Thus, to obtain robustness against imperfect CSI, the CSI imperfection level parameter should be optimized.

\subsubsection{Convexity of the objective function with respect to $\alpha$} 

By using (\ref{eq.gI}), we can expand (\ref{objective}) in terms of the components of \( \mathbf{g}_k \) as follows: 
\begin{equation}\label{obj.2}
    J = \sum_{m=1}^{M} |g_{mk}|^2=\sum_{m=1}^{M} \beta_{mk} f_{mk}\left ( \alpha  \right ),
\end{equation}
where
\begin{equation}
\begin{split}
    f_{mk}\left ( \alpha  \right )=&(1-\alpha)|h_{mk}|^2 + \alpha|\tilde{h}_{mk}|^2 + \\
    &\sqrt{\alpha(1-\alpha)}(h_{mk}\tilde{h}_{mk}^* + h_{mk}^*\tilde{h}_{mk})=\\
    &(1-\alpha)|h_{mk}|^2 + \alpha|\tilde{h}_{mk}|^2 + \\
    &\sqrt{\alpha(1-\alpha)}2\text{Re}\{h_{mk}\tilde{h}_{mk}^*\}.
\end{split}
\end{equation}
Taking the first derivative with respect to $\alpha$, we obtain
\begin{equation}\label{der.J}
\begin{split}
    \frac{\partial f_{mk}\left ( \alpha  \right )}{\partial \alpha}=&-|h_{mk}|^2+|\tilde{h}_{mk}|^2+\\
    &\frac{1-2\alpha }{\sqrt{a\left ( 1-\alpha  \right )}}2\text{Re}\{h_{mk}\tilde{h}_{mk}^*\}.
\end{split}
\end{equation}
Thus, the second derivative with respect to $\alpha$ is obtained as
\begin{equation}
    \frac{\partial^{2} f_{mk}\left ( \alpha  \right )}{\partial \alpha^{2}}=\frac{-\text{Re}\{h_{mk}\tilde{h}_{mk}^*\}}{\alpha \left ( 1-\alpha  \right )\sqrt{\alpha \left ( 1-\alpha  \right )}}.
\end{equation}
Accordingly, we obtain
\begin{equation} \label{dev.w.r.s.alpha}
    \frac{\partial J}{\partial \alpha}=\sum_{m=1}^{M} \beta_{mk}q_{{mk}}\left ( \alpha  \right ),
\end{equation}
where $q_{{mk}}\left ( \alpha  \right )$ is given as
\begin{equation}
    q_{{mk}}\left ( \alpha  \right )= -|h_{mk}|^2+|\tilde{h}_{mk}|^2+\frac{1-2\alpha }{\sqrt{a\left ( 1-\alpha  \right )}}2\text{Re}\{h_{mk}\tilde{h}_{mk}^*\}. 
\end{equation}
Then the second derivative of the objective function with respect to \(\alpha \) is
\begin{equation}
    \frac{\partial^{2} J}{\partial \alpha^{2}}=-\sum_{m=1}^{M} \beta_{mk}\frac{\text{Re}\{h_{mk}\tilde{h}_{mk}^*\}}{\alpha \left ( 1-\alpha  \right )\sqrt{\alpha \left ( 1-\alpha  \right )}}=r\left ( \alpha  \right ).
\end{equation}
To discuss the convexity of the function $J$, we need to consider the sign of the second derivative of the function $f_{mk}\left ( \alpha  \right )$. Given that \( h_{mk} \) and \( \tilde{h}_{mk} \) are independent Gaussian random variables representing small-scale fading, their product \( h_{mk}\tilde{h}_{mk}^* \) and their real part, \(\text{Re}\{h_{mk}\tilde{h}_{mk}^*\}\), are also random. This randomness affects the sign of the second derivative and, therefore, the convexity of \( J \). The denominator \(\alpha(1-\alpha)\sqrt{\alpha(1-\alpha)}\) is always positive for \(\alpha \in (0,1)\), as it is a product and square root of positive terms within this interval. The sign of the numerator, \(-\text{Re}\{h_{mk}\tilde{h}_{mk}^*\}\), depends on the real part of the product of the random variables which could be positive or negative. 

\subsubsection{Problem Solution}
\begin{itemize}
    \item $r\left ( \alpha  \right )> 0$

    In this case, the objective function $J$ would be convex for all \(\alpha \in (0,1)\). 
\begin{itemize}
\item{Max-min problems}: 
For solution of the max-min problems such as what we have in (\ref{rop2}) and (\ref{rop4}), where $J$ is the objective function defined as (\ref{obj.2}), we utilize a gradient descent technique where the channel estimation error parameter $\alpha$ is updated as 
\begin{equation}
    \alpha \left ( i \right )=\alpha \left ( i -1\right )-\mu \frac{\partial J}{\partial \alpha }\bigg|_{\alpha = \alpha \left ( i -1\right )},
\end{equation}
where $i$ is the iteration index, $\mu$ is the positive step size and $\frac{\partial J}{\partial \alpha }$ is obtained using (\ref{dev.w.r.s.alpha}). Next, we will project the result to the feasible region defined by $\beta\leq \left\| \tilde{\mathbf{g}}_{k}\right\|^{2}\leq \beta_0$ such that if the constraint is not satisfied, $\alpha$ is updated. In order to update $\alpha$, we first rewrite $\tilde{\mathbf{g}}_{k}$ in (\ref{chan.vec.k}) as
\begin{equation}
    \tilde{\mathbf{g}}_{k}=\sqrt{\alpha }\left [ \sqrt{\beta _{1k}}\tilde{h}_{1k},\cdots , \sqrt{\beta _{Mk}}\tilde{h}_{Mk}   \right ]^{T}=\sqrt{\alpha }\mathbf{\Gamma}_{k}.
\end{equation}
Then, $\alpha$ is updated as
\begin{align} \label{projection}
        \alpha \left ( i \right )= 
        \begin{cases}
        \frac{\beta}{\| \mathbf{\Gamma}_{k} \|^{2}}, & \text{if } \alpha < \frac{\beta}{\| \mathbf{\Gamma}_{k} \|^{2}} \\
        \frac{\beta_0}{\| \mathbf{\Gamma}_{k} \|^{2}}, & \text{if } \alpha > \frac{\beta_0}{\| \mathbf{\Gamma}_{k} \|^{2}} 
        \end{cases}
    \end{align}
    After computing $\alpha$ and the objective function for every UE, the UE for which the objective function is maximized is selected as the solution to  (\ref{rop2}).
    \item Min-max problems:
    In order to solve the min-max problems such as that in (\ref{rop3}) using the objective function given in (\ref{obj.2}) is convex, similar to the explanations in Section \ref{WRGDPA}, we evaluate the objective function at the constraint boundaries $\|\tilde{\mathbf{g}}_{k}\|^{2} = \beta$ and $\|\tilde{\mathbf{g}}_{k}\|^{2} = \beta_0$. Thus, $\alpha$ is obtained by
\begin{align}
        \alpha = 
        \begin{cases}
        \frac{\beta}{\| \mathbf{\Gamma}_{k} \|^{2}}, & \text{if } \ J \bigg|_{\alpha=\frac{\beta}{\| \mathbf{\Gamma}_{k} \|^{2}}} > J \bigg|_{\alpha=\frac{\beta_0}{\| \mathbf{\Gamma}_{k} \|^{2}}} \\
        \frac{\beta_0}{\| \mathbf{\Gamma}_{k} \|^{2}}, & \text{if }  \ J \bigg|_{\alpha=\frac{\beta_0}{\| \mathbf{\Gamma}_{k} \|^{2}}} > J \bigg|_{\alpha=\frac{\beta}{\| \mathbf{\Gamma}_{k} \|^{2}}} 
        \end{cases}
    \end{align}
    After finding $\alpha$ for UEs in the set $\mathcal{S}_{n}(j)$, we select the UE which minimizes the objective function as the solution to (\ref{rop3}).
 \end{itemize}
 \item $r\left ( \alpha  \right )< 0$ \label{min-max.p.s}

 In this case, the objective $J$ is concave for all \(\alpha \in (0,1)\). 
 
\begin{itemize}
    \item Max-min problems: To solve the minimization in the problems such as (\ref{rop2}) and (\ref{rop4}), since the objective is concave, the global minimum will occur at the boundary of the feasible region as shown in Fig.~\ref{figconcave}, where the concave function $f\left ( \textup{x} \right )$ is examined within the interval $\textup{x}_{1}\leq \textup{x}\leq \textup{x}_{2}$, and its minimum value occurs at the endpoint $\textup{x}_{1}$. 
     \begin{figure}
    \centering
    \includegraphics[width=.725\linewidth]{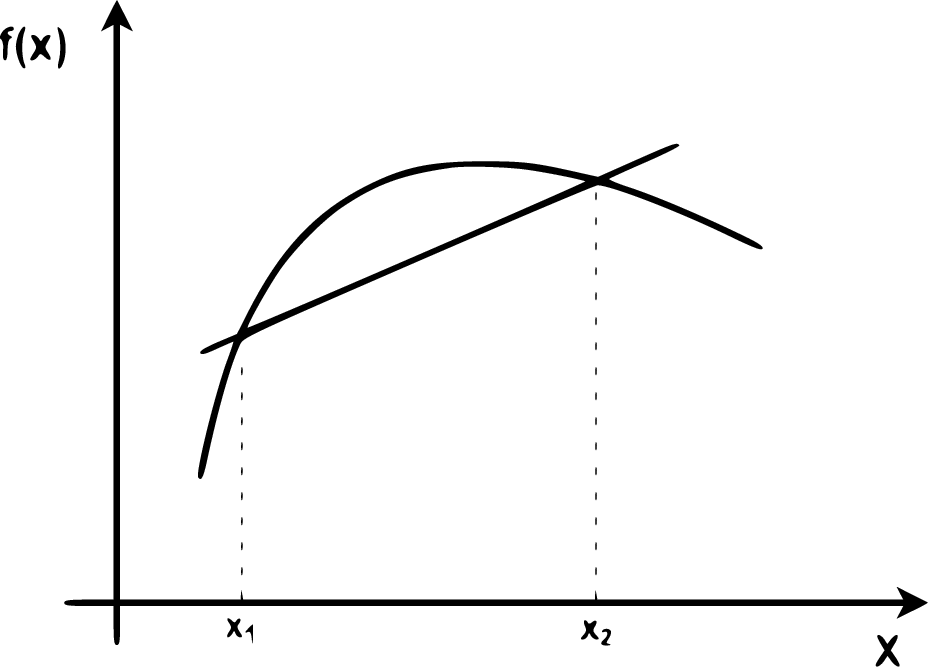}
    \vspace{-1.25em}
    \caption{Boundary extremum illustration, concave case.}
    \label{figconcave}
\end{figure}
    Thus, $\alpha$ is obtained as
    \begin{align}
        \alpha = 
        \begin{cases}
        \frac{\beta}{\| \mathbf{\Gamma}_{k} \|^{2}}, & \text{if } \ J \bigg|_{\alpha=\frac{\beta}{\| \mathbf{\Gamma}_{k} \|^{2}}} < J \bigg|_{\alpha=\frac{\beta_0}{\| \mathbf{\Gamma}_{k} \|^{2}}} \\
        \frac{\beta_0}{\| \mathbf{\Gamma}_{k} \|^{2}}, & \text{if }  \ J \bigg|_{\alpha=\frac{\beta_0}{\| \mathbf{\Gamma}_{k} \|^{2}}} < J \bigg|_{\alpha=\frac{\beta}{\| \mathbf{\Gamma}_{k} \|^{2}}} 
        \end{cases}
    \end{align}
    Given $\alpha$ for every UE, the one with the highest objective is the desired UE.
    \item Min-max problems: To solve the maximizationthe problems such as (\ref{rop3}), we use a gradient ascent technique in which the channel estimation error parameter is updated using the following equation:
    \begin{equation}
    \alpha \left ( j \right )=\alpha \left ( j -1\right )+\eta \frac{\partial J}{\partial \alpha }\bigg|_{\alpha = \alpha \left ( j -1\right )},
\end{equation}
where $j$ is the iteration index, $\eta$ is the positive step size and $\frac{\partial J}{\partial \alpha }$ is obtained using (\ref{dev.w.r.s.alpha}). Subsequently, if the obtained parameter does not satisfy the constraint $\beta\leq \left\| \tilde{\mathbf{g}}_{k}\right\|^{2}\leq \beta_0$, we will project the result to the feasible region using the same method as in (\ref{projection}).

After obtaining $\alpha$ for every UE in the given set of UEs, the UE with the smallest objective $J$ is selected as the solution to min-max problems.

\end{itemize}
\end{itemize}

\subsection{Convexity Analysis of the Conditioned Error Expectation} \label{Analys.Er}

For convexity analysis of the objective function with respect to the power allocation factors, in (\ref{exp.err.con}), the first derivative is obtained as (\ref{first.or}) as described in Section \ref{RPA1}.
 To take the second derivative with respect to $\mathbf{d}$, the only focus is on the term $2\rho _{f}( \mathbf{W}^{H}\hat{\mathbf{G}}_{a}^*\hat{\mathbf{G}}_{a}^T\mathbf{W}\textup{diag} ( \mathbf{d} ) )\odot \mathbf{I}$ which includes the power allocation matrix.  Since $\mathbf{W}^{H}\hat{\mathbf{G}}_{a}^*\hat{\mathbf{G}}_{a}^T\mathbf{W}$ is positive semi-definite and elements of $\mathbf{d}$ are positive, we can consider the $\mathbf{W}^{H}\hat{\mathbf{G}}_{a}^*\hat{\mathbf{G}}_{a}^T\mathbf{W}\textup{diag}( \mathbf{d} )$ as positive semi-definite. Equation (\ref{first.or}) keeps the main diagonal of a positive semi-definite matrix which are non-negative. In matrix derivative rules, the derivative of a matrix with respect to a matrix as $\frac{d\mathbf{A}}{d\mathbf{B}}$, we should calculate derivative of $\mathbf{A}$ with respect to every element of $\mathbf{B}$ each one results in a matrix. Thus, if we take the second derivative of the objective function  with respect to $\mathbf{D}$, it will result in a set of $n$ matrices. Using (\ref{der2}) shows that each matrix will be a matrix with one non-negative element on the main diagonal and other elements are zero which is also positive semi-definite. Thus, we conclude that the objective $\mathbb{E} [ \varepsilon \mid \hat{\mathbf{G}}_a ]$ is convex with respect to the power allocation matrix. 

\subsection{{Analysis of the error 
with respect to 
$\tilde{\mathbf{G}}_a$}} \label{Analys.Er.Ga}

{Since $\mathbb{E}[\epsilon]$ given in \eqref{exp.err} incorporates the expectation $\mathbb{E}_{\tilde{\mathbf{G}}_a}$ over the channel estimation errors, the objective function no longer explicitly depends on the instantaneous realization of $\tilde{\mathbf{G}}_a$. As a result, any derivatives of the expected objective $\mathbb{E}[\epsilon]$ with respect to $\tilde{\mathbf{G}}_a$ become zero, i.e., $\frac{\partial \mathbb{E}[\epsilon]}{\partial \tilde{\mathbf{G}}_a} = 0$, and thus there is no direct way to optimize $\mathbb{E}[\epsilon]$ with respect to $\tilde{\mathbf{G}}_a$. Therefore, to address imperfect CSI, we introduce and optimize the parameter $\alpha$, treating it as a deterministic design variable that can be tuned to improve performance under channel uncertainty.}

\subsection{Computational complexity}

For complexity comparison of the C-ESG and RC-ESG, we note that for calculation of the $\mathbf{g}_{{k}}^{H}\mathbf{g}_{{k}}$ of all UEs, the required complexity is  $\mathcal{O}( KNL )$. For sum-rate calculation, considering the matrix inversion based on Equation (\ref{eq:RCF}), we consider the complexity $\mathcal{O} ( (NL  )^{3}  )$, and considering scheduled UEs, we need $\mathcal{O} ( n (NL   )^{3}  )$ calculations. Thus, for C-ESG, the complexity is $\mathcal{O} ( n\left (NL  \right )^{3}  )+\mathcal{O} ( KNL )$. For RC-ESG, considering $\textup{I}_{R}$ iterations in solution of the REOP optimization problems and the same complexity for sum-rate expressions, the complexity is  $\mathcal{O} ( n (NL   )^{3}  )+\mathcal{O} ( \textup{I}_{R}KNL  )$. Depending on the number of iterations $\textup{I}_{R}$, the complexity of the RC-ESG, is larger than C-ESG due to the inclusion of channel estimation errors and worst-case optimization. For GDPA and RGDPA, the complexities are of orders $\mathcal{O} ( 3 (NL )^{2}n\textup{I}_{D}  )$ and $\mathcal{O} ( 4 (NL  )^{2}n\textup{I}_{D}  )$, respectively. The additional complexity in the RGDPA algorithm arises from computing the extra term specified in (\ref{first.or}), as compared to (\ref{der1.err.exp}). For WRGDPA, calculation of (\ref{Second,dev}) imposes the complexity of order $\mathcal{O} (24 (NL  )^{2}n  )$. Next, we need $\mathcal{O} (9 (NL  )^{2}n\textup{I}_{G}  )$ calculations to obtain the desired $\alpha$ using (\ref{First,dev}), and $\mathcal{O} ( 3 (NL   )^{2}n\textup{I}_{D}  )$ calculations to obtain desired power allocation factors. Thus, WRGDPA has complexity  $\mathcal{O} (24  (NL  \ )^{2}n  )+\mathcal{O} (9 (NL  )^{2}n\textup{I}_{G}  )+ \mathcal{O} ( 3 (NL   )^{2}n\textup{I}_{D}  )$.

\section{Simulation Results} \label{Simul} 

In this section, we consider a user centric cell-free network including $L=16$ APs each equipped with $N=4$ antennas, $K=32$ UEs, and the C-ESG multiuser scheduling technique to select $n=16$ UEs. The network area, path loss and shadowing parameters are based on the network used in \cite{mashdour2022enhanced}.We consider the systems with perfect CSI and imperfect CSI with various levels of CSI uncertainty. In all cases of robust resource allocation, we set the channel estimation error power bounds (including $\beta$, $\beta_0$, $\beta_1$ and $\beta_2$)  such that a range of $0.05 \leq \alpha \leq 0.3$ results for imperfect channel factor. Note that this interval is determined to ensure the channel remains within a balanced range - neither deviating significantly towards an unreliable state nor approaching too closely perfect CSI.  {In all simulation results, the sum-rate is averaged over multiple independent channel realizations to mitigate the randomness and ensure reliable performance evaluation.} 

\begin{figure}
	\centering
		\includegraphics[width=.8\linewidth]{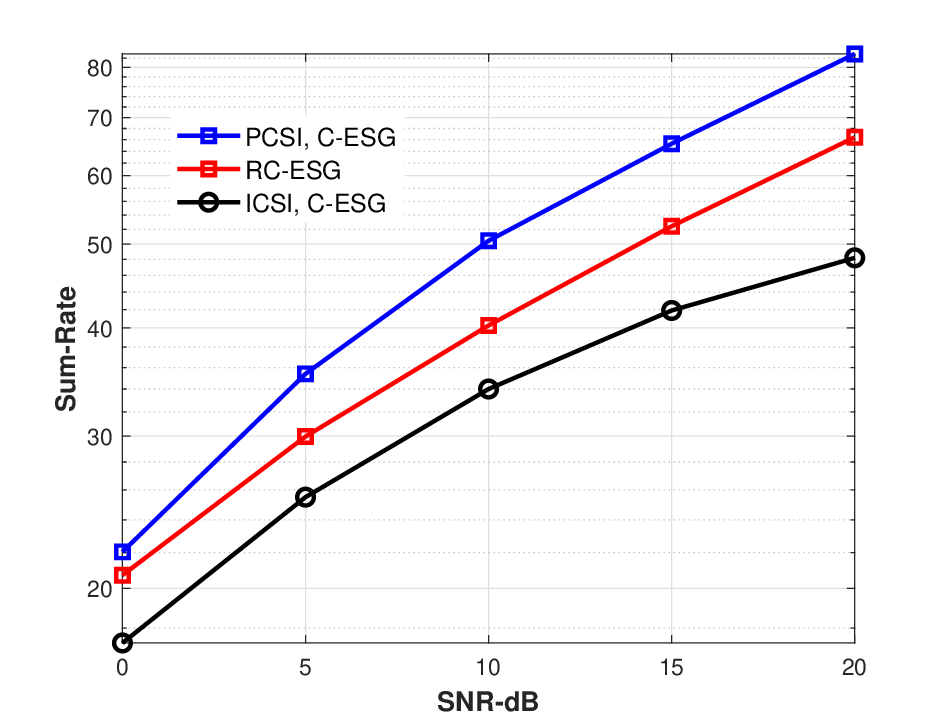}
      \vspace{-1em}
	\caption{\small{Comparison of the C-ESG multiuser scheduling in perfect CSI and imperfect CSI UCCF networks and the RC-ESG for MMSE precoder and EPL power loading, when $\alpha=0.15$ for imperfect CSI case, $L=16$, $N=4$, $K=32$ and $n=16$.}}
	\label{fig:fig1}
\end{figure}

The proposed RC-ESG is compared with C-ESG under perfect CSI (PCSI) and imperfect CSI (ICSI), all with EPL. We set $\alpha = 0.15$ for the ICSI network, and the results are shown in Fig.~\ref{fig:fig1}. It is evident that RC-ESG outperforms C-ESG in the context of ICSI, approaching the performance seen in networks with PCSI. {Quantitatively, the RC-ESG improves the sum-rate by approximately 10\%-20\% across different SNR levels, highlighting its robustness in handling imperfect CSI.}

\begin{figure}
	\centering
		\includegraphics[width=.8\linewidth]{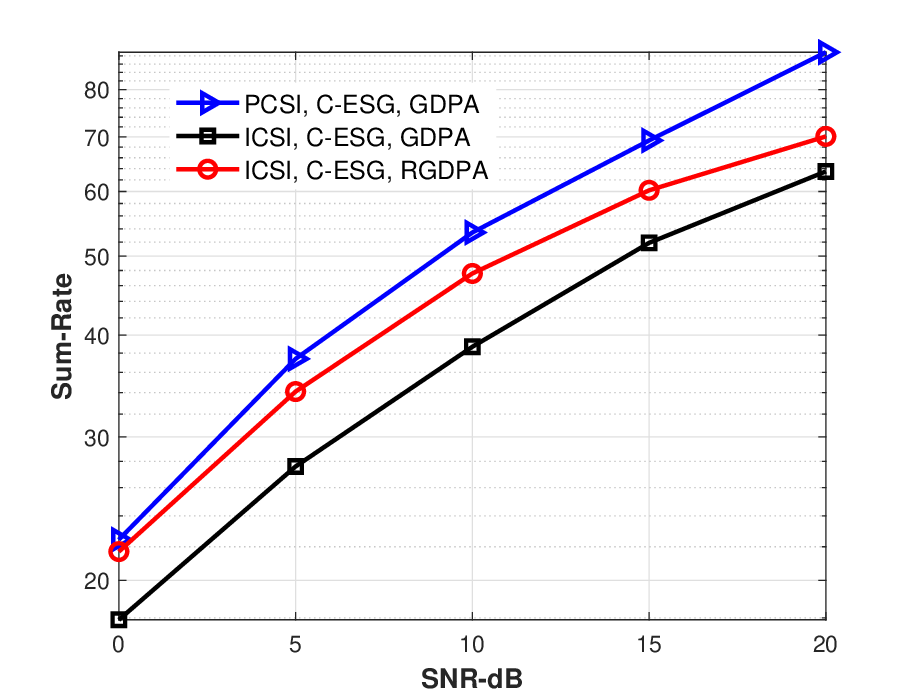}
      \vspace{-1em}
	\caption{\small{Comparison of the RGDPA power allocation, GDPA with ICSI and GDPA with PSCI, when $\alpha=0.15$ for imperfect CSI case, $L=16$, $N=4$, $K=32$, $n=16$ and MMSE precoder.}}
	\label{fig:fig6}
\end{figure}

In Fig.~\ref{fig:fig6}, we compared the proposed GDPA and RGDPA power allocation algorithms where we used the C-ESG scheduling algorithm. {It is evident that RGDPA provides better performance, with a sum-rate increase of 15\%-25\% compared to GDPA under ICSI, and approaches the performance of PCSI.} The uncertainty level was set to $\alpha=0.15$ for ICSI.

\begin{figure}
	\centering
		\includegraphics[width=.8\linewidth]{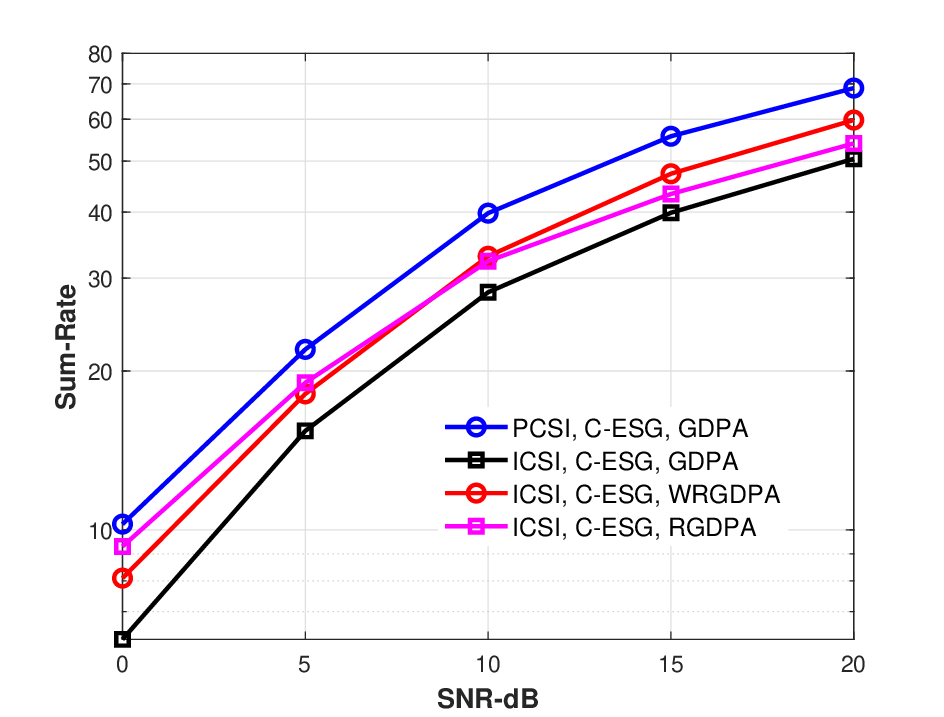}
      \vspace{-1em}
	\caption{\small{Comparison of WRGDPA and RGDPA power allocation, GDPA with ICSI and GDPA with PSCI, when $\alpha=0.15$ for imperfect CSI case, $L=16$, $N=4$, $K=32$, $n=16$ and ZF precoder.}}
	\label{fig:fig7}
\end{figure}

In Fig.~\ref{fig:fig7}, we consider ZF precoding for the network and after user scheduling using C-ESG, the proposed WRGDPA and RGDPA power allocation techniques are compared with the GDPA power allocation for both PCSI and ICSI when the imperfect CSI factor $\alpha=0.15$ is considered in ICSI. The results sshowthat WRGDPA and RGDPA have significantly outperformed GDPA with ICSI and approached the PCSI case. A comparison of WRGDPA and RGDPA shows that RGDPA has better performance for lower SNRs, whereas as SNR increases WRGDPA outperforms RGDPA.

\begin{figure}
	\centering
		\includegraphics[width=.8\linewidth]{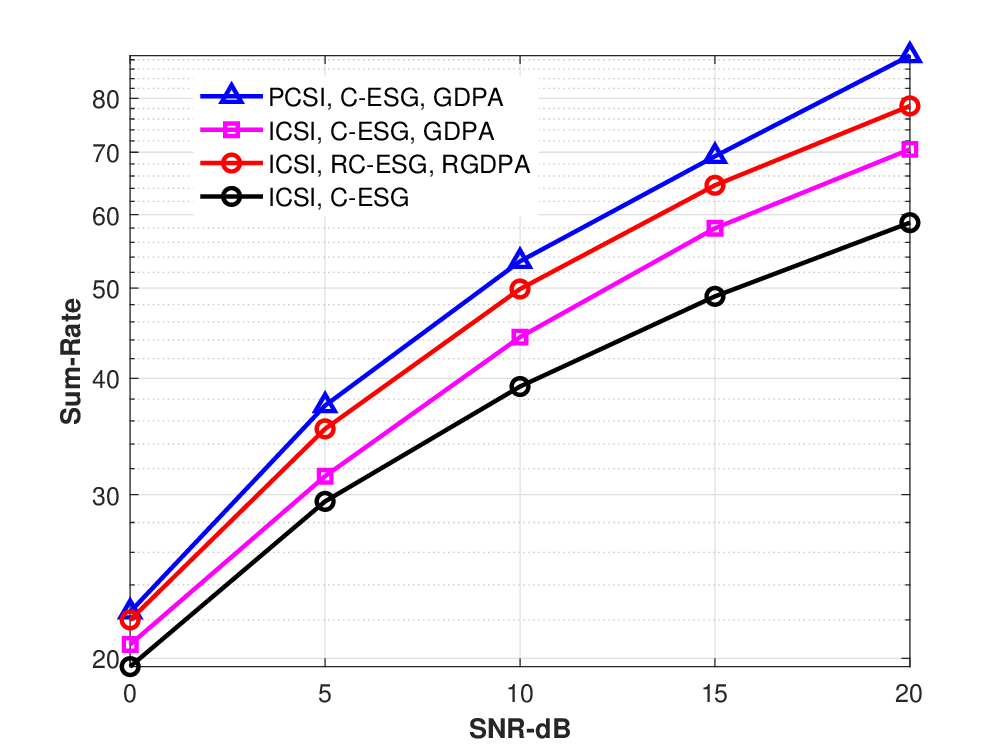}
      \vspace{-1em}
      \caption{\small{Performance of the proposed resource allocation technique, when $\alpha=0.1$ for imperfect CSI case, $L=16$, $N=4$, $K=32$, $n=16$ and MMSE precoder.}}
	\label{fig:fig8}
\end{figure}

In Fig.~\ref{fig:fig8}, we combined RC-ESG with RGDPA to assess the performance of the proposed robust resource allocation techniques. {In this case, the imperfection level for the ICSI scenario has been reduced from $\alpha=0.15$ to $\alpha=0.1$. This adjustment demonstrates the effect of reduced CSI imperfection on the performance of the proposed resource allocation techniques. The results show that even with different levels of imperfection, the proposed techniques consistently yield a sum-rate improvement of up to 20\% compared to existing approaches under ICSI, further validating the effectiveness of the proposed techniques.}

\section{Conclusions} \label{conclud}
 In this paper, we
introduced a robust resource allocation framework that significantly improves the performance of cell-free networks in the presence of imperfect CSI. Our approach employs a sequential strategy with a robust user scheduling algorithm based on worst-case optimization and two distinct power allocation strategies: one according to the worst-case robust optimization and another based on conditional MSE minimization. {These approaches have been analyzed analytically and their effectiveness is demonstrated through simulation results, which show a sum-rate improvement of up to 30\% under various SNR levels and CSI conditions, approaching the performance of systems with perfect CSI. These findings confirm that our proposed framework effectively mitigates the adverse effects of imperfect CSI and leads to substantial enhancements in network performance.}
 \appendices
 \section{Sum-rate expression derivation}
  \label{sum-rate}
Considering the received signal $\mathbf{y}_{a}$ in UCCF networks in (\ref{UCCF.sig}), we write the terms for channel error and noise as $\mathbf{y}_{a_{e,w}}=\sqrt{\rho _{f}}\tilde{\mathbf{G}}_{a}^{T}\mathbf{P}_{a}\mathbf{x}+\mathbf{w}
$. 
{Then, an upper bound on the achievable sum-rate under imperfect channel knowledge is given by}
\begin{equation}
 C_{ch}= \textup{log}_{2}\left ( \textup{det}\left ( \pi e \mathbf{R}_{\mathbf{y}_{a}} \right ) \right )-\textup{log}_{2}\left ( \textup{det}\left ( \pi e \mathbf{R}_{\mathbf{y}_{a_{e,w}}} \right ) \right ),  
\end{equation}
where
\begin{equation}
\mathbf{R}_{\mathbf{y}_a}=\mathbb{E}\left [ \mathbf{y}_{a}\mathbf{y}_{a}^{H} \right ],
\end{equation}
and 
\begin{equation}
\mathbf{R}_{\mathbf{y}_{a_{e,w}}}=\mathbb{E}\left [ \mathbf{y}_{a_{e,w}}\mathbf{y}_{a_{e,w}}^{H} \right ], 
\end{equation}
{where it has been assumed knowledge of the channel error and the noise distribution.}
 
{We expand the inner terms in $\mathbf{R}_{\mathbf{y}_a}$, which results in}
\begin{equation}
    \begin{split}
        {\mathbf{R}_{\mathbf{y}_a}=}&{E\left[\rho _{f}\tilde{\mathbf{G}}_{a}^{T}\mathbf{P}_{a}\mathbf{x}\mathbf{x}^{H}\mathbf{P}_{a}^{H}\tilde{\mathbf{G}}_{a}^{*}\right]+E\left[\sqrt{\rho _{f}}\tilde{\mathbf{G}}_{a}^{T}\mathbf{P}_{a}\mathbf{x}\mathbf{w}^{H}\right]+}\\
    &{E\left[\sqrt{\rho _{f}}\mathbf{w}\mathbf{x}^{H}\mathbf{P}_{a}^{H}\tilde{\mathbf{G}}_{a}^{*}\right]+E\left[\mathbf{w}\mathbf{w}^{H}\right]+}\\
    &{E\left[\rho _{f}\hat{\mathbf{G}}_{a}^{T}\mathbf{P}_{a}\mathbf{x}\mathbf{x}^{H}\mathbf{P}_{a}^{H}\hat{\mathbf{G}}_{a}^{*}\right]+}\\
    &{E\left[\rho _{f}\hat{\mathbf{G}}_{a}^{T}\mathbf{P}_{a}\mathbf{x}\mathbf{x}^{H}\mathbf{P}_{a}^{H}\tilde{\mathbf{G}}_{a}^{*}\right]+E\left[\sqrt{\rho _{f}}\hat{\mathbf{G}}_{a}^{T}\mathbf{P}_{a}\mathbf{x}\mathbf{w}^{H}\right]+}\\
    &{E\left[\rho _{f}\tilde{\mathbf{G}}_{a}^{T}\mathbf{P}_{a}\mathbf{x}\mathbf{x}^{H}\mathbf{P}_{a}^{H}\hat{\mathbf{G}}_{a}^{*}\right]+E\left[\sqrt{\rho _{f}}\mathbf{w}\mathbf{x}^{H}\mathbf{P}_{a}^{H}\hat{\mathbf{G}}_{a}^{*}\right]} 
    \end{split}
\end{equation}
{Since we consider \(\hat{\mathbf{G}}_a\) as constant within a coherence interval but \(\tilde{\mathbf{G}}_a\) as stochastic with zero mean,  independent and uncorrelated with $\mathbf{x}$ and $\mathbf{w}$, and $\mathbf{P}_{a}$ as a function of $\hat{\mathbf{G}}_{a}$ only, we can rewrite $\mathbf{R}_{\mathbf{y}_a}$ as follows:}
\begin{equation} \label{Expand.Ry}
    \begin{split}
        {\mathbf{R}_{\mathbf{y}_a}=}&{\rho_f \mathbf{P}_a \mathbb{E}\left[\tilde{\mathbf{G}}_a^T \tilde{\mathbf{G}}_a^*\right] \mathbb{E}\left[\mathbf{x} \mathbf{x}^H\right] \mathbf{P}_a^H+}\\
        &{\sqrt{\rho_f} \mathbb{E}\left[\tilde{\mathbf{G}}_a^T\right] \mathbf{P}_a \mathbb{E}\left[\mathbf{x} \mathbf{w}^H\right]+}\\
    &{\sqrt{\rho_f} \mathbb{E}\left[\mathbf{w}\mathbf{x}^H\right] \mathbf{P}_a^H \mathbb{E}\left[\tilde{\mathbf{G}}_a^*\right]+E\left[\mathbf{w}\mathbf{w}^{H}\right]+}\\
    &{\rho_f \hat{\mathbf{G}}_a^T \mathbf{P}_a \mathbb{E}\left[\mathbf{x} \mathbf{x}^H\right] \mathbf{P}_a^H \hat{\mathbf{G}}_a^*+}\\
    &{\rho_f \hat{\mathbf{G}}_a^T \mathbf{P}_a \mathbb{E}\left[\mathbf{x} \mathbf{x}^H\right] \mathbf{P}_a^H \mathbb{E}\left[\tilde{\mathbf{G}}_a^*\right]+\sqrt{\rho_f} \hat{\mathbf{G}}_a^T \mathbf{P}_a \mathbb{E}\left[\mathbf{x} \mathbf{w}^H\right]+}\\
    &{\rho_f \mathbb{E}\left[\tilde{\mathbf{G}}_a^T\right] \mathbf{P}_a \mathbb{E}\left[\mathbf{x} \mathbf{x}^H\right] \mathbf{P}_a^H \hat{\mathbf{G}}_a^*+\sqrt{\rho_f} \mathbb{E}\left[\mathbf{w} \mathbf{x}^H\right] \mathbf{P}_a^H \hat{\mathbf{G}}_a^*} 
    \end{split}
\end{equation}
{In \eqref{Expand.Ry}, $\mathbb{E}[\mathbf{x}\mathbf{x}^H] = \mathbf{I}_K$, $\mathbb{E}[\mathbf{w}\mathbf{w}^H] = \sigma_w^2 \mathbf{I}_K$, $\mathbb{E}\left[\mathbf{x} \mathbf{w}^H\right]=\mathbb{E}\left[\mathbf{w} \mathbf{x}^H\right]=0$ and $\mathbb{E}\left[\tilde{\mathbf{G}}_a^T\right]=\mathbb{E}\left[\tilde{\mathbf{G}}_a\right]=0$. Thus, the terms with $\mathbb{E}\left[\mathbf{x} \mathbf{w}^H\right]$ and $\mathbb{E}\left[\tilde{\mathbf{G}}_a\right]$ go to zero and (79) simplifies to}
\begin{equation}
    \begin{split}
        {\mathbf{R}_{\mathbf{y}_a}=}&{\rho_f \mathbf{P}_a \mathbb{E}\left[\tilde{\mathbf{G}}_a^T \tilde{\mathbf{G}}_a^*\right] \mathbf{P}_a^H+}{\sigma_w^2 \mathbf{I}_K+}{\rho_f \hat{\mathbf{G}}_a^T \mathbf{P}_a \mathbf{P}_a^H \hat{\mathbf{G}}_a^*}
    \end{split}
\end{equation}
{Since \( \mathbf{P}_a \) and \( \mathbf{P}_a^H \) are deterministic matrices, we can move them outside of the expectation. The expectation operation only acts on the random matrix \( \tilde{\mathbf{G}}_a \). Thus, we have:}
\begin{equation}
    \begin{split}
        {\mathbf{R}_{\mathbf{y}_a}=}&{\rho _{f}\mathbb{E}\left[\tilde{\mathbf{G}}_a^T \mathbf{P}_a \mathbf{P}_a^H \tilde{\mathbf{G}}_a^*\right]+}{\sigma_w^2 \mathbf{I}_K+}{\rho_f \hat{\mathbf{G}}_a^T \mathbf{P}_a \mathbf{P}_a^H \hat{\mathbf{G}}_a^*}. \label{Rya}
    \end{split}
\end{equation}

{Since $\mathbf{y}_{a_{e,w}}=\sqrt{\rho _{f}}\tilde{\mathbf{G}}_{a}^{T}\mathbf{P}_{a}\mathbf{x}+\mathbf{w}$, we obtain $\mathbf{R}_{\mathbf{y}_{a_{e,w}}}$ as follows:} 
\begin{equation} \label{yaeyae}
    \begin{split}
    {\mathbf{R}_{\mathbf{y}_{a_{e,w}}}=}&{E\left[ \mathbf{y}_{a_{e,w}}\mathbf{y}_{a_{e,w}}^{H}\right] =}\\
    &{E\left[ \left ( \sqrt{\rho _{f}}\tilde{\mathbf{G}}_{a}^{T}\mathbf{P}_{a}\mathbf{x}+\mathbf{w} \right )\left ( \sqrt{\rho _{f}}\tilde{\mathbf{G}}_{a}^{T}\mathbf{P}_{a}\mathbf{x}+\mathbf{w} \right )^{H}\right]=}\\
    &{E\left[\rho _{f}\tilde{\mathbf{G}}_{a}^{T}\mathbf{P}_{a}\mathbf{x}\mathbf{x}^{H}\mathbf{P}_{a}^{H}\tilde{\mathbf{G}}_{a}^{*}\right]+E\left[\sqrt{\rho _{f}}\tilde{\mathbf{G}}_{a}^{T}\mathbf{P}_{a}\mathbf{x}\mathbf{w}^{H}\right]+}\\
    &{E\left[\sqrt{\rho _{f}}\mathbf{w}\mathbf{x}^{H}\mathbf{P}_{a}^{H}\tilde{\mathbf{G}}_{a}^{*}\right]+E\left[\mathbf{w}\mathbf{w}^{H}\right]} 
    \end{split}
\end{equation}
{Then, \eqref{yaeyae} is rewritten as}

    \begin{equation}
        \begin{split}
            {\mathbf{R}_{\mathbf{y}_{a_{e,w}}}=}&{\rho_f \mathbf{P}_a \mathbb{E}\left[\tilde{\mathbf{G}}_a^T \tilde{\mathbf{G}}_a^*\right] \mathbb{E}\left[\mathbf{x} \mathbf{x}^H\right] \mathbf{P}_a^H+}\\
            &{\sqrt{\rho_f} \mathbb{E}\left[\tilde{\mathbf{G}}_a^T\right] \mathbf{P}_a \mathbb{E}\left[\mathbf{x} \mathbf{w}^H\right]+}\\
    &{\sqrt{\rho_f} \mathbb{E}\left[\mathbf{w}\mathbf{x}^H\right] \mathbf{P}_a^H \mathbb{E}\left[\tilde{\mathbf{G}}_a^*\right]+E\left[\mathbf{w}\mathbf{w}^{H}\right]=}\\
    &{\rho_f \mathbf{P}_a \mathbb{E}\left[\tilde{\mathbf{G}}_a^T \tilde{\mathbf{G}}_a^*\right] \mathbf{P}_a^H+}{\sigma_w^2 \mathbf{I}_K} 
        \end{split}
    \end{equation}
    {Now, moving \( \mathbf{P}_a \) and \( \mathbf{P}_a^H \) into the expectation operation, we obtain}
\begin{equation}\label{Ryaew}
    {\mathbf{R}_{\mathbf{y}_{a_{e,w}}}=\rho _{f}\mathbb{E}\left[\tilde{\mathbf{G}}_a^T \mathbf{P}_a \mathbf{P}_a^H \tilde{\mathbf{G}}_a^*\right]+\sigma_{w}^{2}\mathbf{I}_{K}.} 
\end{equation}

Now, we substitute (\ref{Rya}) and (\ref{Ryaew}) in the following equation
\begin{equation}\label{Cch}
\begin{split}
  &\log\left ( \det\left ( \pi e \mathbf{R}_{\mathbf{y}_{a}} \right ) \right )-\log\left ( \det\left ( \pi e \mathbf{R}_{\mathbf{y}_{a_{e,w}}} \right ) \right )=\\
  &\log\left ( \frac{\det\left ( \pi e \mathbf{R}_{\mathbf{y}_{a}}  \right )}{\det\left ( \pi e \mathbf{R}_{\mathbf{y}_{a_{e,w}}} \right )} \right )=\log \left ( \det \left ( \mathbf{R}_{\mathbf{y}_{a}} \mathbf{R}_{\mathbf{y}_{a_{e,w}}}^{-1} \right ) \right ) 
  \end{split}
\end{equation}
Eq. (\ref{Cch}) simplifies to the sum-rate expression given in (\ref{eq:RCF}).

  \section{Convexity of the objective $J$ with respect to small scale fading error vector}
  \label{der2toh}
  The objective function \( J \) is defined as 
\begin{equation}
    J =\sum_{m=1}^{M} g_{mk}g_{mk}^* =\sum_{m=1}^{M} \beta_{mk}f_{mk},
\end{equation}
\begin{equation} 
\begin{split}
    f_{mk}=
    &\left[(1-\alpha)h_{mk}h_{mk}^* + \alpha\tilde{h}_{mk}\tilde{h}_{mk}^* +\right. \\
    &\left.\sqrt{\alpha(1-\alpha)}(h_{mk}\tilde{h}_{mk}^* + h_{mk}^*\tilde{h}_{mk})\right]
\end{split}
\end{equation}
 Using Equation (\ref{der2}), we obtain
\begin{equation}
    \frac{\partial f_{mk}}{\partial \tilde{h}_{mk}}=\alpha\tilde{h}_{mk}^*+\sqrt{\alpha(1-\alpha)}h_{mk}^*
\end{equation}

Thus, the first and the second derivatives with respect to $\tilde{h}_{mk}$ are given as follows, respectively,
\begin{equation}
    \frac{\partial J}{\partial \tilde{\mathbf{h}}_{k}}=\begin{bmatrix}
\beta_{1k}\left ( \alpha\tilde{h}_{1k}^*+\sqrt{\alpha(1-\alpha)}h_{1k}^* \right )\\ 
\vdots \\ 
\beta_{Mk}\left ( \alpha\tilde{h}_{Mk}^*+\sqrt{\alpha(1-\alpha)}h_{Mk}^* \right )
\end{bmatrix}
\end{equation}
Since $\frac{\partial^{2} f_{mk}}{\partial \tilde{h}_{mk}^{2}}=0$, the second derivative is obtained as
\begin{equation}
    \frac{\partial^{2} J}{\partial \tilde{\mathbf{h}}_{k}^{2}}=\mathbf{0}.
\end{equation}
which shows that the 
function is affine with respect to $\tilde{\mathbf{h}}_{k}$.

\section{ Error expression derivation}
  \label{Error Expression}

Using (\ref{yc2}) and (\ref{err.def}), the error equation is expanded as
 
\begin{equation} \label{eq:err}
  \begin{split}
      \varepsilon &=\mathbf{x}^{H}\mathbf{x}+\mathbf{w}^{H}\mathbf{w}+\\
      &\rho_{f}\mathbf{x}^{H}\textup{diag}\left ( \mathbf{d}\right)\mathbf{W}^{H}\hat{\mathbf{G}}_{a}^*\hat{\mathbf{G}}_{a}^T\mathbf{W}\textup{diag}\left ( \mathbf{d}\right)\mathbf{x}+\\
&\rho_{f}\mathbf{x}^{H}\mathbf{P}_{a}^{H}\tilde{\mathbf{G}}_{a}^*\tilde{\mathbf{G}}_{a}^T\mathbf{P}_{a}\mathbf{x}-\\
&\sqrt{\rho _{f}}\mathbf{x}^{H}\hat{\mathbf{G}}_{a}^T\mathbf{W}\textup{diag}\left ( \mathbf{d}\right)\mathbf{x}-\sqrt{\rho_{f}}\mathbf{x}^{H}\textup{diag}\left ( \mathbf{d}\right)\mathbf{W}^{H}\hat{\mathbf{G}}_{a}^*\mathbf{x}-\\
&\sqrt{\rho _{f}}\mathbf{x}^H {\tilde{\mathbf{G}}}_{a}^T\mathbf{P}_{a}\mathbf{x}-\sqrt{\rho _{f}}\mathbf{x}^{H}\mathbf{P}_{a}^{H}\tilde{\mathbf{G}}_{a}^*\mathbf{x}-\\
&\mathbf{x}^{H}\mathbf{w}-\mathbf{w}^{H}\mathbf{x}+\rho_{f}\mathbf{x}^{H}\textup{diag}\left ( \mathbf{d}\right)\mathbf{W}^{H}\hat{\mathbf{G}}_{a}^*{\tilde{\mathbf{G}}}_{a}^T\mathbf{P}_{a}\mathbf{x}+\\
&\rho_{f}\mathbf{x}^{H}\mathbf{P}_{a}^{H}\tilde{\mathbf{G}}_{a}^*\hat{\mathbf{G}}_{a}^T\mathbf{W}\textup{diag}\left ( \mathbf{d}\right)\mathbf{x}+\\
&\sqrt{\rho _{f}}\mathbf{x}^{H}\textup{diag}\left (\mathbf{d}\right)\mathbf{W}^{H}\hat{\mathbf{G}}_{a}^*\mathbf{w}+\sqrt{\rho_{f}}\mathbf{w}^{H}\hat{\mathbf{G}}_{a}^T\mathbf{W}\textup{diag}\left ( \mathbf{d}\right )\mathbf{x}+\\
&\sqrt{\rho _{f}}\mathbf{w}^{H}\tilde{\mathbf{G}}_{a}^T\mathbf{P}_{a}\mathbf{x}+\sqrt{\rho _{f}}\mathbf{x}^{H}\mathbf{P}_{a}^{H}\tilde{\mathbf{G}}_{a}^*\mathbf{w}
\end{split}
\end{equation}

Since the trace operator retains a scalar unchanged, using the property ${Tr}\left ( \mathbf{A}+\mathbf{B} \right )={Tr}\left ( \mathbf{A} \right )+{Tr}\left ( \mathbf{B} \right )$, where $\mathbf{A}$ and $\mathbf{B}$ are matrices of same dimensions, the error is rewritten as 
\begin{equation} \label{eq:trerr}
 \begin{split}
          \varepsilon & ={Tr}\left (\mathbf{x}^{H}\mathbf{x}  \right )- {Tr}\left(\mathbf{x}^{H}\mathbf{w}  \right)-{Tr}\left (\mathbf{w}^{H}\mathbf{x}  \right )+{Tr}\left (\mathbf{w}^{H}\mathbf{w}  \right )+\\ 
       &{Tr}\left (\rho _{f}\mathbf{x}^{H}\textup{diag}\left ( \mathbf{d}\right )\mathbf{W}^{H}\hat{\mathbf{G}}_{a}^*\hat{\mathbf{G}}_{a}^T\mathbf{W}\textup{diag}\left ( \mathbf{d}\right )\mathbf{x}  \right )+\\
       &{Tr}\left (\rho_{f}\mathbf{x}^{H}\mathbf{P}_{a}^{H}\tilde{\mathbf{G}}_{a}^*\tilde{\mathbf{G}}_{a}^T\mathbf{P}_{a}\mathbf{x}\right )-\\
       &{Tr}\left (\sqrt{\rho _{f}}\mathbf{x}^{H}\hat{\mathbf{G}}_{a}^T\mathbf{W}\textup{diag}\left ( \mathbf{d}\right )\mathbf{x}  \right )-\\
       & {Tr}\left (\sqrt{\rho _{f}}\mathbf{x}^{H}\textup{diag}\left ( \mathbf{d}\right )\mathbf{W}^{H}\hat{\mathbf{G}}_{a}^*\mathbf{x}  \right )-\\
       &{Tr}\left ( \sqrt{\rho _{f}}\mathbf{x}^H {\tilde{\mathbf{G}}}_{a}^T\mathbf{P}_{a}\mathbf{x} \right )-{Tr}\left ( \sqrt{\rho _{f}}\mathbf{x}^{H}\mathbf{P}_{a}^{H}\tilde{\mathbf{G}}_{a}^*\mathbf{x} \right )+\\
       &{Tr}\left ( \rho_{f}\mathbf{x}^{H}\textup{diag}\left ( \mathbf{d}\right)\mathbf{W}^{H}\hat{\mathbf{G}}_{a}^*{\tilde{\mathbf{G}}}_{a}^T\mathbf{P}_{a}\mathbf{x}\right )+\\
   &{Tr}\left ( \rho_{f}\mathbf{x}^{H}\mathbf{P}_{a}^{H}\tilde{\mathbf{G}}_{a}^*\hat{\mathbf{G}}_{a}^T\mathbf{W}\textup{diag}\left ( \mathbf{d}\right)\mathbf{x} \right )+\\
       & {Tr}\left (\sqrt{\rho _{f}}\mathbf{x}^{H}\textup{diag}\left ( \mathbf{d}\right )\mathbf{W}^{H}\hat{\mathbf{G}}_{a}^*\mathbf{w}  \right )+\\
       &{Tr}\left (\sqrt{\rho _{f}}\mathbf{w}^{H}\hat{\mathbf{G}}_{a}^T\mathbf{W}\textup{diag}\left ( \mathbf{d}\right )\mathbf{x}  \right )+\\
       &{Tr}\left ( \sqrt{\rho _{f}}\mathbf{w}^{H}\tilde{\mathbf{G}}_{a}^T\mathbf{P}_{a}\mathbf{x} \right )+{Tr}\left ( \sqrt{\rho _{f}}\mathbf{x}^{H}\mathbf{P}_{a}^{H}\tilde{\mathbf{G}}_{a}^*\mathbf{w} \right ).
\end{split}
\end{equation}

\bibliographystyle{IEEEbib}
\bibliography{refs}

\end{document}